\documentclass[pre,twocolumn,showpacs,amsmath,amssymb,dvipdfm]{revtex4}
\usepackage{graphicx,amstext}% Include figure files
\usepackage{dcolumn}% Align table columns on decimal point
\usepackage{bm}% bold math
\newcommand{\vs}{{\it vs.\@}}
\newcommand{\ie}{{\it i.e.\@}}
\newcommand{\al}{{\it et al.\@}}
\newcommand{\bq}{\begin{equation}}
\newcommand{\eq}{\end{equation}}
\begin{document}
\title{Dielectric relaxation of DNA aqueous solutions}

\author{S.~Tomi\'{c}}
 \homepage{http://real-science.ifs.hr/}
\email{stomic@ifs.hr}
\author{S.~Dolanski Babi\'{c}}
\altaffiliation {Permanent address: Department of physics and biophysics, 
Medical School, University of Zagreb, 10000 Zagreb, Croatia.}
\author{T.~Vuleti\'{c}}
\affiliation{Institut za fiziku, 10000 Zagreb, Croatia}

\author{S.~Kr\v{c}a}
\author{D.~Ivankovi\'{c}}
\affiliation{Institute Rudjer 
Bo\v{s}kovi\'{c}, 10000 Zagreb, Croatia}

\author{L.~Gripari\'{c}}
\affiliation{Department of Biological Chemistry, David Geffen School of 
Medicine, UCLA, Los Angeles, California 90095, USA}

\author{R.~Podgornik}
\affiliation{Department of Physics, University of Ljubljana and J.~Stefan 
Institute, 1000 Ljubljana, Slovenia and \\
Laboratory of Physical and Structural Biology NICHD, National Institutes of Health, Bethesda, Maryland 20892, USA}

\date{\today}

\begin{abstract}

We report on a detailed characterization of complex dielectric response of 
Na-DNA aqueous solutions by means of low-frequency dielectric spectroscopy 
(40 Hz - 110 MHz). Results reveal two broad relaxation modes of strength 
$20<\Delta\varepsilon_{\textrm {LF}}<100$ and $5<\Delta\varepsilon_{\textrm 
{HF}}<20$, centered at 0.5 kHz$<\nu_{\textrm {LF}}<$70 kHz and 0.1 
MHz$<\nu_{\textrm {HF}}<$15 MHz.  The characteristic length scale of the LF 
process, $50<L_{\textrm {LF}}<750$nm, scales with DNA concentration as 
$c_{\textrm{DNA}}^{-0.29\pm0.04}$ and is independent of the ionic strength in 
the low added salt regime. Conversely, the measured length scale of the LF  
process does not vary with DNA  concentration but depends on the ionic strength of 
the added salt as $I_s^{-1}$ in the high added salt regime. On  the other hand, the  
characteristic length scale of the HF process, $3<L_{\textrm {HF}} <50$~nm,  
varyes  with DNA concentration as $c_{\textrm{DNA}}^{-0.5}$ for intermediate 
and large DNA concentrations. At low DNA concentrations and  in the low added 
salt limit the  characteristic length scale of the HF process scales as  
$c_{\textrm{DNA}}^{-0.33}$. We put these results in perspective regarding the 
integrity of the double stranded form of DNA at low salt conditions as well as 
regarding the role of different types of  counterions in different regimes of dielectric 
dispersion. We argue that the free DNA  counterions are primarily active  in the HF 
relaxation,  while the condensed  counterions play a role only in the LF relaxation. 
We also suggest theoretical interpretations  for all these length scales in the whole 
regime of DNA and salt concentrations and discuss their ramifications and limitations. 

\end{abstract}

\pacs{{82.39.Pj }{87.15.He }{77.22.Gm }}

\maketitle

\section{Introduction}
\label{sec1}
Semiflexible polyelectrolytes are ubiquitous in biological context  
ranging from charged biopolymers such as
DNA or filamentous $F$ actin, and then all the way to molecular  
aggregates such as bacterial $fd$ viruses or the tobacco mosaic  
virus. They are an essential and fundamental component of the  
cellular environment and make their mark in its every structural and  
functional aspect \cite{Daune03}. Their role is not confined solely  
to various (macro) molecular assemblies in the biological {\sl  
milieu} but are equally prevalent in colloidal systems and soft  
matter in general \cite{Osawa71,Schmitz93}. It is  their  
connectivity, stiffness and strong electrostatic interactions, that  
allow polyelectrolytes to show a wide range of complex behaviors,  
depending on their concentration, their overall length and the  
concentration and valency of the added salt ions and intrinsic  
counterions \cite{Dobrynin}.

In the simplest case of monovalent counterions, polyelectrolytes are usually 
stretched due to electrostatic repulsions and therefore statistically assume a 
rod-like configuration \cite{Netz03}. Polyvalent counterions on the other hand 
can turn electrostatic  repulsion into attraction \cite{Boroudjerdi05}. This 
{\em s.c.} correlation effect is one of the most important features of the 
polyvalent counterions and has fundamental repercussions for all charged soft 
matter. Correlation attraction can strongly reduce the rigidity of charged 
polymers which then collapse into highly compact states \cite{Hansen99}.

Deoxyribonucleic acid (DNA) is in many respects a paradigm of a stiff, highly 
charged polymer. The structural origin of the charge is due to negatively 
charged phosphate groups positioned along the DNA backbone \cite{Grosberg02}. The 
nominal charge density of double stranded DNA amounts to one negative elementary 
charge per 1.7 \AA~or two elementary charges per base pair. When dealing with 
electrostatic interactions and their consequences in DNA one can usually ignore 
the internal chemical structure of DNA except as it transpires through the bare value of the persistence length (see below). 

Electrostatic interactions in stiff polyelectrolytes with monovalent counterions 
and added salt in aqueous solutions are standardly approached {\em  via} the 
Poisson - Boltzmann (PB) theory that combines electrostatics with statistical 
mechanics on a simplified mean-field level \cite{Schmitz93}. On the PB level the 
outcome of the competition between entropy of mobile charges and their 
electrostatic interaction energy leads {\em  grosso modo} to two types of 
effects.

The first one is connected with electrostatic interaction of counterions and 
fixed charges on the polyelectrolyte. Positively charged counterions are 
attracted to the surface of the negatively charged DNA where they tend to 
accumulate in a layer of condensed counterions. This accumulation can be 
understood within the framework of the Manning-Oosawa (MO) counterion 
condensation theory \cite{Schmitz93} as well as on the level of the solutions of 
the PB equation \cite{Hansen01}. Within this theory the counterions accumulate 
in the condensed layer around the cylindrical DNA surface only if the charge 
density parameter $\eta > 1$. Here $\eta= z l_B / b$, where $z$ is the valency 
of the counterion, $b$ is the linear charge spacing and $l_B$ is the Bjerrum 
length, defined as

\begin{equation}
l_B = {\textrm e}_0^2/(4 \pi \varepsilon \varepsilon_0 kT)
\label{Bjerrumlength}
\end{equation}

Here ${\textrm e}_0$ is elementary charge, $\varepsilon_0$ is permeability of vacuum, $kT$ is thermal energy scale and $\varepsilon$ is the dielectric constant of the solvent. Bjerrum length is obviously defined as the separation between charges at which  their electrostatic interaction energy equals their thermal energy. In aqueous 
solutions $l_B=7.1$~\AA. The charge density parameter $\eta$ thus measures the 
relative strength of electrostatic interactions \vs{} thermal motion and is 
strongly dependent on the valency of the counterions. In the MO theory the 
counterions accumulate in the condensed layer exactly to such an extent that the 
effective charge density parameter $\eta$ is reduced to 1 \cite{Bloomfield00}, 
\ie{} the effective separation between charges is increased from $b$ to $l_B$. 
The condensed counterions in the MO theory are still assumed to be perfectly 
mobile.  Because of counterion condensation the effective charge of DNA is reduced by a 
factor $r = 1 - 1/(z\eta)$. It is important to note here that the concept of 
counterion condensation is intrinsically nonlinear and is a fundamental property 
of highly charged polymers. This includes DNA in both its double stranded 
($b=1.7$~\AA, $\eta = 4.2$) as well as single stranded ($b=4.3$~\AA, $\eta = 
1.7$) forms. Counterion condensation for monovalent DNA salts is experimentally 
observed directly by small-angle X-ray scattering measurements \cite{Das03}. For 
double stranded DNA and monovalent counterions, counterion condensation occurs 
with $r = 0.76$. In the MO counterion condensation theory the condensation 
occurs only if the salt concentration is low enough to satisfy $\kappa^{-1} \gg 
a$, where $a$ is the polymer radius and $\kappa^{-1}$ is the Debye screening 
length (see below) \cite{Bloomfield00}. Furthermore the counterion condensation 
strictly occurs only in the limit of vanishingly small concentration of DNA 
\cite{Hansen01}. For finite DNA concentrations a more complicated model of 
counterion condensation has to be invoked \cite{Hansen01} that is based on the 
solution of the PB equation in the cell model geometry. 

The second effect of competition between entropy of mobile charges and their 
electrostatic interaction energy is due to the interaction of mobile ions in 
solution between themselves and their redistribution in the field of fixed 
polyelectrolyte charges. This redistribution leads to screening of electrostatic 
interactions between fixed charges. For small enough, fixed, charge density the 
screening is accurately described by the Debye-H\"{u}ckel equation, which is 
just a linearized form of the PB equation \cite{Schmitz93}. On the 
Debye-H\"{u}ckel level the screening is quantified by a screening or Debye length 
$\kappa^{-1}$ defined as

\begin{equation}
\kappa^{2}=8 \pi l_B n
\label{Debyelength}
\end{equation}
	
where again $l_B$ is the Bjerrum length and n is the density of added salt. For 
monovalent salts the Debye length in \AA~is given numerically as  $3.04 I_s^{-1/2}$, where $I_s$  is the ionic strength in M. Indeed for monovalent salts both effects, ionic screening as well as counterion condensation, coexist. In general, however, one can not invoke screening effects in polyvalent salt solutions since 
in that case the whole PB conceptual framework breaks down and correlation, not 
screening, effects \cite{Boroudjerdi05} become the salient feature of the 
behavior of the system. The ionic screening in monovalent salt solutions, 
augmented by the effect of thermal DNA undulations, has been shown to 
quantitatively  describe the measured osmotic pressure of DNA solutions in a 
fairly wide range of concentrations \cite{Strey99,Podgornik89,Podgornik94}.

Apart from being a charged polymer, DNA is also molecularly rather stiff. The 
flexibility of polymers is usually described {\sl via} the persistence length $L_p$. 
Persistence length is nothing but the correlation length for orientational 
correlations along the molecular axis of the polymer \cite{Daune03}. For DNA the 
usually accepted value is about 500 \AA~\cite{Bloomfield00}. Persistence length 
separates two regimes of behavior: rigid chain regime for contour lengths 
smaller than $L_p$ and flexible chain regime for contour lengths much larger then $L_p$. The persistence length depends on long range electrostatic 
interactions along the polyelectrolyte chain. The influence of electrostatic 
interactions on the persistence length was first considered by Odijk 
\cite{Odijk77} and independently by Skolnick and Fixman \cite{Skolnick77}. 
According to the Odijk-Skolnick-Fixman (OSF) theory, the total persistence 
length can be decomposed into a structural ($L_0$) and electrostatic ($L_e$) 
contribution as

\begin{equation}
L_p = L_0 + L_e = L_0  +  l_B /(2 b \kappa)^{2} 
\label{Persistlength}
\end{equation}
						
where $b$ is again the separation between charges. Assuming the MO condensation, 
\ie{} $b= l_B$, one gets $L_p = L_0 + 0.324 I_s^{-1}$ in \AA. As is clear from 
the OSF theory, counterion condensation reduces the electrostatic contribution 
to the persistence length due to an increase in the effective separation between 
charges from $b$ to $l_B$. The OSF result, though it can be nominally applied 
only at restrictive conditions, appears to work well when compared to 
experiments \cite{Baumann97} as well as computer simulations \cite{Ullner03}. On the other hand, in the regime of no added salt and thus weak electrostatic screening due to other chains and counterions, a semiflexible charged chain in a semidilute polyelectrolyte solution behaves like a random walk of correlation blobs (see below) with chain size $R \propto c^{-0.25}$ \cite{Dobrynin}. Coming back to the electrostatic persistence length, it is worth mentioning that $L_e$ strongly depends on the valency of the counterions and in general on the details of the electrostatic interaction potential. For monovalent counterions 
$L_e$ is usually positive indicating an effective repulsion between monomers. 

It is evident that starting assumptions  $\kappa^{-1} \gg a$ of the MO theory 
(infinitely thin polymers or highly dilute polyelectrolyte solutions) and 
$\kappa^{-1} \gg b$ of the OSF theory (monovalent salt with low enough 
screening) enforce their limited validity. It is not clear at all whether one can 
apply these theories in the limit where added salt concentration is smaller than 
the DNA concentration, {\em viz.} the concentration of intrinsic counterions. 
This fact raises further questions, even in the simplest case of DNA in 
monovalent salt solutions, of the general applicability of these models to 
predict the amount of counterion condensation, as well as to describe properties 
like fundamental length scales of DNA in solution. 

Due to the intrinsic length of native DNA, the concentration of  polyelectrolyte 
solutions that we are dealing with in this contribution is always higher than 
the chain overlap concentration $c$*  \cite{Dobrynin}. This means that we are 
effectively always in the semidilute regime, where it has been known for a long 
time \cite{Dobrynin,Odijk79}, that a new length scale emerges describing the density 
correlations in the polyelectrolyte solution. This length scale is equal to the 
correlation length or the mesh size, describing the correlation volume in 
semidilute polyelectrolyte solutions, and  is given by the de Gennes-Pfeuty-Dobrynin (dGPD) correlation length \cite{Dobrynin,deGennes76,Pfeuty78} that 
scales universally with the concentration of the polyelectrolyte, $c$, as  

\begin{equation}
\xi \propto c^{-0.5}
\label{xi05}
\end{equation}

It is interesting that this form of scaling, first derived for uncharged polymer 
solutions, is preserved also in the case of polyelectrolytes with some 
theoretically possible adjustments only in the case of added salt effects, 
and is expected to be proportional to the screening length due to both free DNA 
counterions and added salt ions \cite{Dobrynin}. For polyelectrolyte solutions the  interpretation of this scaling result is that for volumes  smaller than $\xi^3$  the polyelectrolyte chain is stiffened by  electrostatic interactions \cite{Schmidt89}, whereas for scales larger than  $\xi$, it behaves as a free flight chain. 

In an attempt to clarify several of the above issues, we have undertaken an 
investigation of dielectric relaxation properties of DNA solutions that covers a 
broad range of DNA concentrations, as well as added salt concentrations. We have 
used low-frequency dielectric spectroscopy  technique, widely established as a 
direct and non-destructive tool to probe charged entities and their structure in 
various bio-macromolecular systems \cite{Pethig79,Nandi00,Colby}. Our aim was 
not only to verify the predictions of the various theories in the well defined 
conditions, but also to investigate the behavior of DNA at extreme conditions 
like very low and very high added salt limit.  A brief report of this investigation has been published in Ref.~\cite{TomicPRL06}.

Interpretation of our results depends on the nature of the  conformation of DNA in low salt or even pure water  solutions, more specifically on whether DNA is in a single stranded (ss-) or double stranded  (ds-) conformation. We have considered this question very carefully in what follows. Different preparation protocols for DNA solutions  were adopted in order to study the issue of ds-DNA stability in pure water (see  the Materials and Methods section) solutions. Denaturation of DNA was in  particular investigated for pure water DNA solutions. These studies indicate that indeed for semidilute conditions the DNA double-helix was never denatured into two spatially distinguishable and well separated single strands. DNA solutions were additionally characterized by UV spectrophotometry, electrophoresis and atomic emmission spectroscopy.

Our results demonstrate that DNA counterions, free as well as condensed 
in variable proportions, contribute to the oscillating polarization in 
the applied electric field and thus together determine the dielectric response 
of the DNA solution. The characteristic size of the relaxation volume for the 
dielectric response of a semidilute DNA solution is given by one of the three 
fundamental length scales: the dGPD mesh size of the whole polyelectrolyte solution, the OSF salt-dependent persistence length of a single polyelectrolyte chain and the average size of the chain in the salt-free polyelectrolyte solution. The OSF prediction for the persistence length as a function of added salt ionic strength is verified in the high added salt limit giving very good agreement. Our results indicate that by going from the high to the low salt limit and all the way down to the nominally pure water solutions, the characteristic length goes from the persistence length of the DNA in solution to a value that corresponds to the average size of the Gaussian chain composed of correlation blobs which scales as $c_{\textrm{DNA}}^{- 0.25}$. Moreover, an exact condition is established that separates the high from the low added salt regime, which reads $2I_s = 0.4c_{in}$, where $c_{in}$ is the concentration of  DNA counterions, These two regimes differ in whether the added salt ions provide the screening or DNA acts as its own salt. Finally, our results 
confirm the theoretical prediction describing the concentration dependence of 
the mesh size of DNA solutions as $\xi \propto c_{\textrm{DNA}}^{-0.5}$. In the limit of low DNA concentrations and low added salt, the semidilute solution correlation length deviates from the classical polyelectrolyte behavior and follows the scaling $c_{\textrm{DNA}}^{-0.33}$. Possible interpretations of this behavior are discussed and the appearance of locally fluctuating regions with exposed hydrophobic cores is suggested as the mechanism for this anomalous scaling.

\section{Materials and methods}
\label{sec2}	
Salmon testes and calf thymus lyophilized Na-DNA threads were obtained from 
Sigma and Rockland, respectively. Low protein content was declared and verified 
by our UV spectrophotometry measurements, for the former $A_{260} /A_{280} =1.65 - 
1.70$ and for the latter $A_{260} /A_{280} =1.87$.  Inductively coupled plasma - atomic emmission spectroscopy (ICP-AES) 
elemental analysis was performed on the Na-DNA threads dissolved in pure water 
\cite{purewater}. The results have shown that the phosphorus and sodium contents 
were the same, implying that only intrinsic sodium atoms are present. Taking 
into account that for double stranded DNA, monomers correspond to 
base pairs of molecular weight 660 g/mol, this result implies 7\%  by weight  of 
Na$^+$ ions in DNA. This means that the concentration of intrinsic DNA counterions 
and the DNA concentration are related by $c_{in} \left[\textrm{mM}\right] = c_{\textrm{DNA}} \left[\textrm{mg/mL}\right]   \times$ 3 $\mu$mol/mg. Electrophoresis measurements were performed on DNA 
dissolved in pure water and in NaCl electrolyte with different ionic strengths. 
The obtained results have shown consistently the existence of polydisperse DNA fragments, most of them in the range 2 - 20 kbp. Since the scale of 3.4~\AA\ %
corresponds to one base pair, we estimate the range of contour lengths of DNA fragments to be 0.7 - 7 $\mu$m.

DNA solutions with different DNA concentrations and different added salt ionic 
strengths were prepared according to two protocols, which we describe below: 
\begin{itemize}
\item[I] {\em Pure water DNA solutions}: 
Dry DNA threads were dissolved in pure water for 48 hours at 4$^o$C so that the 
solutions within concentration range $0.01 \leq c_{\textrm{DNA}} \leq 15$ mg/mL were 
obtained. The {\em ionic strength} of pure water $I_s \approx 0.01$ mM was 
estimated from the measured conductivity $\sigma = 1.5$ $\mu$S/cm of the pure 
water sample in the chamber for dielectric spectroscopy, using molar 
conductivity of the highly diluted NaCl (126.5 Scm$^{2}$/mol). An increased 
conductivity value of pure water, as compared to the declared one \cite{purewater}, is due to manipulation in  the laboratory environment. 
\item[II] {\em DNA solutions with added salt}: 
\begin{itemize}
\item[II$1$]
NaCl was added to DNA water solution with a chosen $c_{\textrm{DNA}}$ (prepared according to I),  so that the added salt ionic strength was achieved in the 
range 0.01 mM$ \leq I_s \leq 4$~mM. 
\item[II$2$]  
DNA solutions with the same $c_{\textrm{DNA}}$ as in II$1$ and with the added salt ionic strength in the range 1 mM$ \leq I_s \leq 4$~mM were prepared starting from stock DNA 
solutions in which DNA was dissolved in 10 mM NaCl for 48 hours at 4$^o$C. One 
of the stock solutions was dialyzed against 10mM NaCl during 24 hours at 4$^o$C 
(II$2.2$), while the other was not (II$2.1$).  
\item[II$3$] 
DNA solutions with concentrations in the range $0.1 \leq c_{\textrm{DNA}} \leq 
1.25$ mg/mL and with the added salt ionic strength $I_s=$~1 mM were prepared starting from 
a stock DNA solution in which DNA was dissolved in 10 mM NaCl for 48 hours at 4$^o$C. Stock solution was dialyzed against 1 mM NaCl during 24 hours at 4$^o$C.
\end{itemize}
\end{itemize}
 
The stock solutions were stored at 4$^o$C and were diluted just before 
dielectric spectroscopy measurements, which were completed in a few days. UV 
spectrophotometry and electrophoresis measurements were performed within next two weeks. For longer periods the stock solutions were stored at -80$^o$C. pH of all solutions 
were found to be around 7. Similar pH values were measured for DNA solutions 
prepared from 10 mM Tris-EDTA buffer (TE), which adjusts pH to 7.6 (not 
discussed in this article). 

The results of ICP-AES elemental analysis indicated that solutions of Na-DNA 
dissolved in pure water did not contain any additional ions except declared 
sodium ones. This was further confirmed by dielectric spectroscopy measurements, 
which gave identical results for DNA solutions with same DNA concentration and 
added salt ionic strength prepared according to the protocol II$1$ and II$2$.

Spectrophotometry measurements at 260 nm were performed to verify nominal DNA 
concentrations. The concentration was determined assuming an extinction 
coefficient at 260 nm $A_{260} = 20$ for 1 mg/mL, meaning that the measured 
absorption $A_{260} = 1$ corresponds to 0.05 mg/mL \cite{Steiner67,Beers67}. 
Throughout this article we will refer to nominal concentrations, which we have 
found to be in a good agreement with the measured ones. In particular, the 
measured concentrations for DNA solutions II$2$ and II$3$, performed on 10 
aliquots, were consistently smaller by about 20\% than the nominal ones. We 
interpret this difference to be due to water content, not taken into account by the spectrophotometry approach. Indeed, 
lyophilized DNA if kept at 110$^o$C for 30 min loses about 20\% in weight.

Dielectric spectroscopy measurements \cite{Colby} were performed at room temperature (25$^{o}$C) using a set-up which consists of a home-made parallel platinum plate capacitive 
chamber and temperature control unit, in conjunction with the Agilent 4294A 
precision impedance analyzer operating in $\nu= 40$~Hz - 110 MHz frequency 
range. The capacitive chamber enables reliable complex admittance measurements 
with reproducibility of 1.5\% of samples in solution with small volume of 100 
$\mu$L and with conductivities in the range of 1.5 - 2000 $\mu$S/cm. Low ac amplitudes of 50 mV were employed in order to probe the 
DNA response in the linear regime, once we verified that for ac signal levels in 
the range between 20 mV and 500 mV the result was essentially the same.  Lowest 
ac amplitudes were not used in order to avoid extensive averaging. Admittance 
was sampled at 201 frequencies at 27 points per frequency decade. At each 
frequency, admittance was sampled 10 times and averaged.  In addition, three 
consecutive frequency sweeps were taken in order to average out the temperature 
variations. Total time for described measurement amounts to 60 sec. Measurement 
functions are the real part of the admittance $G_{exp}(\omega)$ and the 
capacitance $C_{exp}(\omega)$, where $\omega=2\pi\nu$.

In addition to DNA samples, the reference samples were also measured in order to
minimize stray impedances, including the free ion contribution and electrode
polarization effects, and extract the response due to DNA only
\cite{Saif91,Colby}. Reference samples were chosen as NaCl solutions of
different molarities, adjusted to have the real part of admittances at 100 kHz
the same as DNA solutions. In comparison to these solutions, independent
adjustment of capacitance at 1 kHz demanded the reference NaCl concentrations
that differ at most by 20\%. Since the difference in capacitance between
reference NaCl solutions of similar molarities is approximately constant at
frequencies above the influence of the electrode polarization, adjusting solely
the real part of admittance effectively also adjusts the capacitance up to an
additive term. As a result, the subtraction of the reference response
successfully eliminated influence of spurious effects down to a low frequency
limit in the range 0.5 - 30 kHz, depending on the solution molarity, and up to a
high frequency limit of 30 MHz. Generally speaking, electrode polarization
effects are larger for higher DNA and added salt concentrations, so that the low
frequency bound is shifted to higher frequencies. Assuming that the conductivity
contribution of each entity in the solution is additive \cite{Osawa71}, the DNA
response is given by
$G(\omega)= G_{exp}(\omega) - G_{ref}(\omega)$
and
$C(\omega)= C_{exp}(\omega) - C_{ref}(\omega)$,
where $ G_{ref}(\omega)$, $C_{ref}(\omega)$ is the reference samples response.
Finally, the real and imaginary parts of dieletric function are extracted using
relations
\begin{eqnarray}
\varepsilon^{\prime}(\omega)&=&\frac{l}{S}\frac{C(\omega)}{ \varepsilon_0}\\
\varepsilon^{\prime\prime}(\omega)&=&\frac{l}{S}\frac{G(\omega)}{\omega 
\varepsilon_0}
\label{GB}
\end{eqnarray}

$l/S = 0.1042\pm 0.0008$~cm$^{-1}$ is the chamber constant, where $S = 
0.98$~cm$^{2}$  is the effective electrode cross-section corresponding to the 
sample of 100 $\mu$L and $l = 0.1021\pm 0.0001$~cm is the distance between the
electrodes. The chamber constant was determined by measuring the real part of
admittance at 100 kHz of a 0.01 M and 0.0005 M KCl standard solutions
(Mettler-Toledo). It was also corroborated by the difference in capacitance of
the chamber when empty and with 100 $\mu$L of the pure water. In the latter
case, the dielectric constant of pure water $\epsilon_w=78.65$ was used as a
standard.

Detailed analysis of the DNA response was made in terms of the complex 
dielectric function $\varepsilon(\omega)$ given by a generalization of the Debye 
expression known as the phenomenological Havriliak-Negami (HN) function 
\cite{Havriliak66}
\begin{equation}
\varepsilon(\omega) - \varepsilon_{\textrm{HF}} = \Delta\varepsilon 
\frac{1}{\left(1+(i \omega \tau_{0})^{1-\alpha}\right)^\beta}
\label{HN}
\end{equation}

where $\Delta\varepsilon = \varepsilon_0 - \varepsilon_{\textrm{HF}}$  is the 
strength of the relaxation process, $\varepsilon_0$ is the static dielectric 
constant ($\omega \ll 1/\tau_0$) and $\varepsilon_{\textrm{HF}}$ is the high 
frequency dielectric constant ($\omega \gg 1/\tau_0$). $\tau_0$ is the mean
relaxation time, while $1-\alpha$ and $\beta$ are the shape parameters which
describe the symmetric broadening of the relaxation time distribution function
and skewness, respectively. We worked with $\beta=1$ since this simplified HN
formulation (also known as the Cole-Cole function) has been widely and
successfully used to describe relaxation processes in disordered systems. 

Measured data were analyzed by using the least squares method in the complex 
plane \cite{Pinteric01, Hinrichs97}. Such an approach, which takes into consideration both 
the real and the imaginary part of the dielectric function at the same time, strongly 
improves the resolution if compared with the method in which the real and 
imaginary parts are treated separately. The complex plane method proved itself 
to be a powerful tool to resolve reliably two close modes in frequency even if 
the strength of one or both modes does not exceed $\Delta\varepsilon =2-3$, 
provided that the ratio of their dielectric strengths is smaller than the ratio 
of their positions in frequency.

\section{Results}
\label{sec3}

We now present results of the dielectric response study performed on DNA 
solutions prepared according to protocols described in Section~\ref{sec2}. The obtained dielectric relaxation results for salmon testes and calf thymus DNA samples were essentially the same. For the sake of simplicity, in the remainder of this article we will speak of DNA 
samples only.

%%%%%%%%%%%%%%FIGURE%%%%%%%%%%%%%%%%%%%%%%%%
\begin{figure}
\resizebox{0.33\textwidth}{!}{\includegraphics*{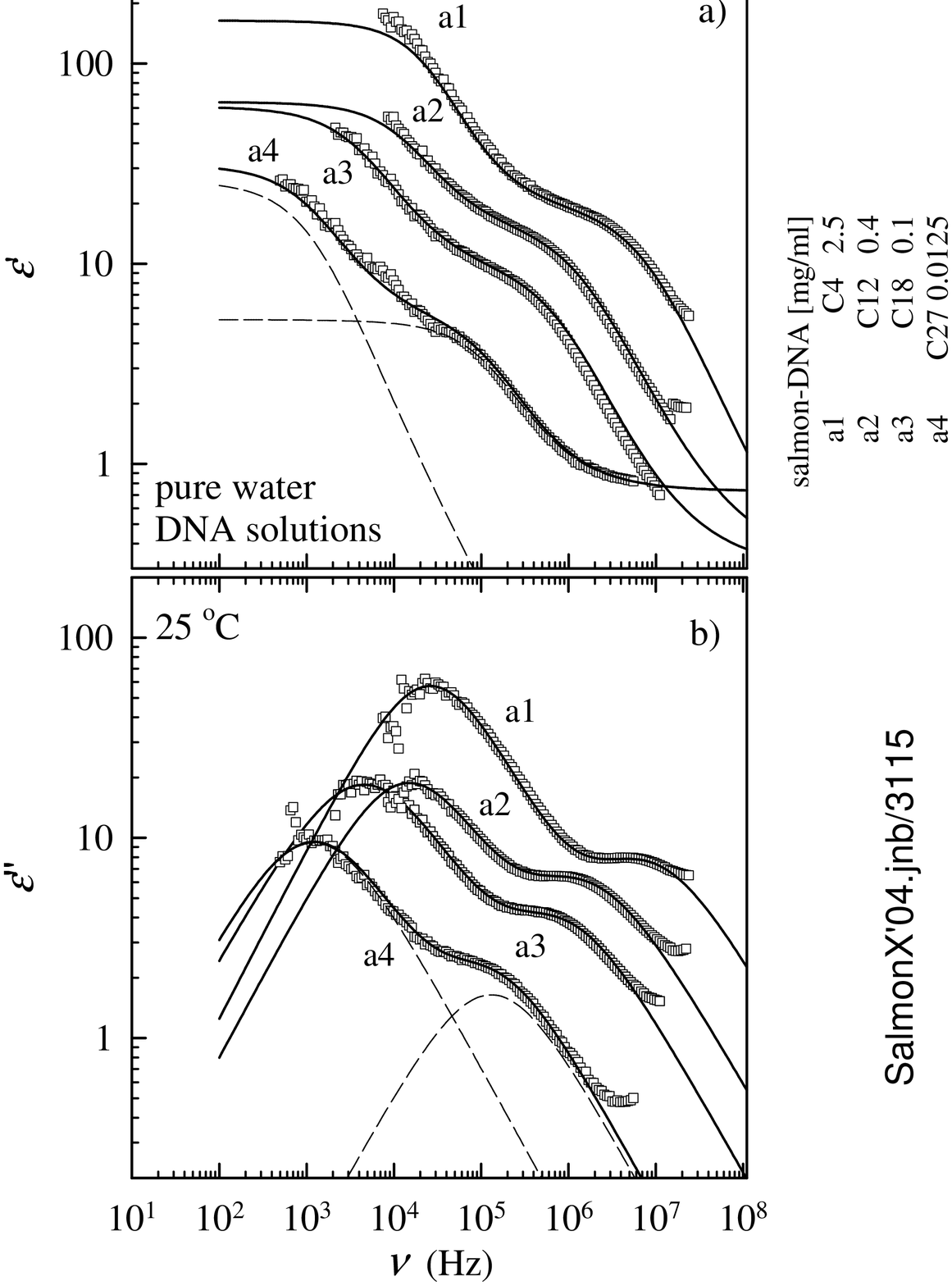}}
\resizebox{0.33\textwidth}{!}{\includegraphics*{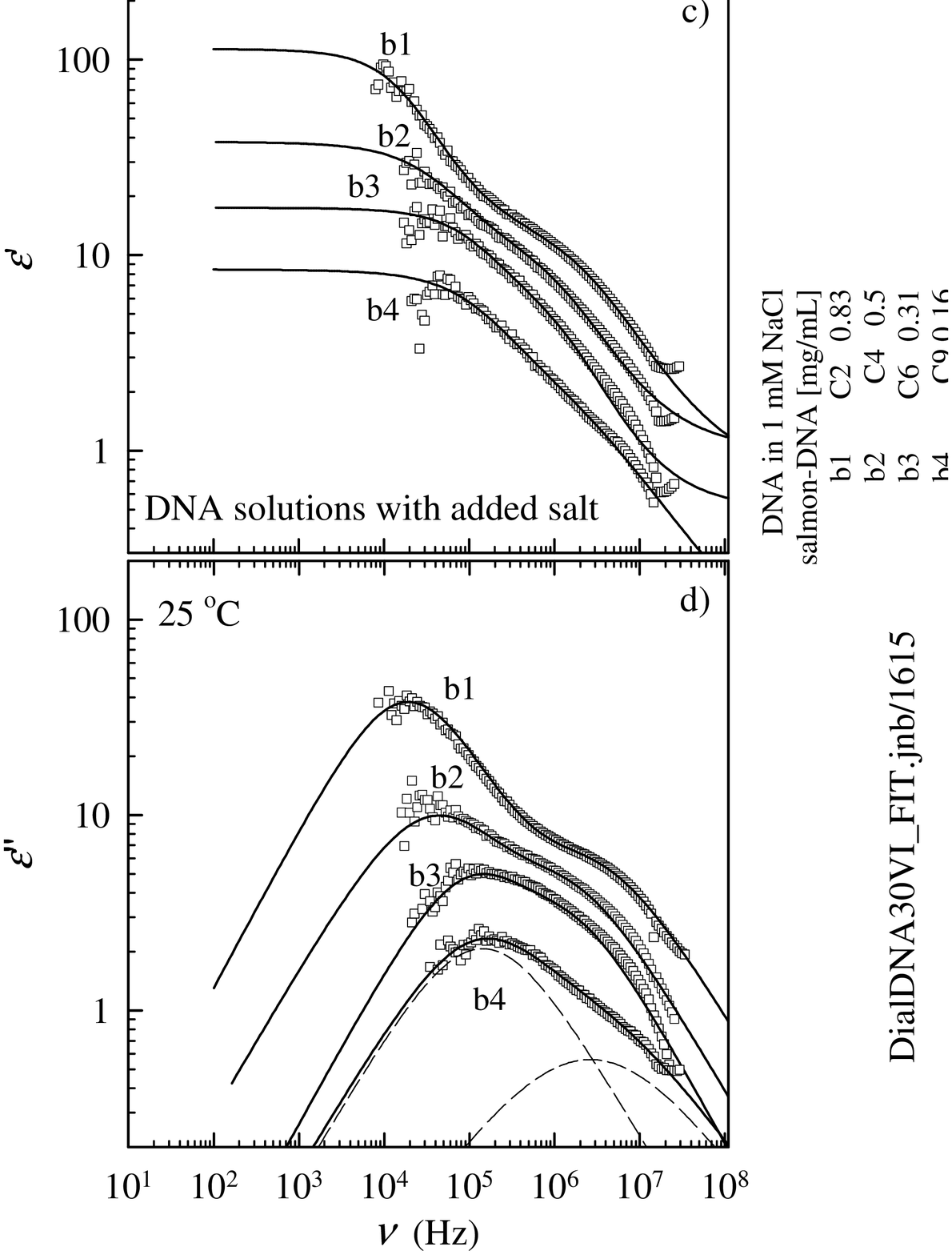}}
%SX'04_Allepsimre3113A.eps S_in_NaClVI'05_Allepsimre1613
\caption{
Double logarithmic plot of the frequency dependence of the real ($\varepsilon$') and imaginary ($\varepsilon$'')
part of the dielectric function at $T=25^{o}$C of (a, b) pure water DNA 
solutions (protocol I) and (c, d) DNA water solutions with added salt $I_s = 1$ 
mM (protocol II$3$) for  representative a1-a4 (2.5, 0.4, 0.1, 0.0125 mg/mL) and 
b1-b4 (0.83, 0.5, 0.31, 0.125 mg/mL) DNA concentrations. The full lines are fits 
to the sum of the two HN forms; the dashed lines represent a single HN form.
}
\label{fig3}
\end{figure}
%%%%%%%%%%%%%%FIGURE%%%%%%%%%%%%%%%%%%%%%%%%

Fig.~\ref{fig3} shows the frequency dependent real and imaginary part of the 
dielectric function for selected DNA concentrations of DNA solutions. The 
results for pure water DNA solutions (protocol I, concentrations a1=2.5 mg/mL, 
a2=0.4 mg/mL, a3=0.1 mg/mL and a4=0.0125 mg/mL) are shown in panel a) and b), 
while the results for DNA solutions with added salt of ionic strength $I_s = 1$ 
mM (protocol II$3$, concentrations b1=0.83 mg/mL, b2=0.5 mg/mL, b3=0.31 mg/mL 
and b4=0.125 mg/mL) are shown in panel c) and d). The observed dielectric 
response is complex and the data were only successfully fitted to a formula 
representing the sum of two HN functions. The full lines in Fig.~\ref{fig3} 
correspond to these fits, while the dashed lines represent single HN forms. The 
main features of this response, for pure water DNA solutions, as well as for DNA 
solutions with added salt, are two broad modes, whose amplitude and position in 
frequency depend on the DNA concentration. The parameter $1 -\alpha$, which 
describes the symmetrical broadening of the relaxation time distribution 
function, is concentration independent and similar for both modes $1 -\alpha 
\approx 0.8$. The mode centered at higher frequencies (0.1 MHz$<\nu_{\textrm 
{HF}}<$15 MHz) is characterized by smaller dielectric strength 
($5<\Delta\varepsilon_{\textrm {HF}}<20$) than the mode 
($20<\Delta\varepsilon_{\textrm {LF}}<100$) centered at lower frequencies (0.5 
kHz$<\nu_{\textrm {LF}}<$70 kHz). In the remainder of this article, we will 
refer to these modes as the high-frequency (HF) and low-frequency (LF) mode, 
respectively.  

In what follows, we discuss possible assignments for these relaxation modes 
inside the framework of existing theoretical approaches for polarization 
response of charged biopolymers in solution 
\cite{Colby,Mandel77,Mandel84,Dukhin80,OBrian86}. An applied ac field generates an oscillating flow of net charge associated with DNA counterions \cite{counterions}
and induces polarization. Since the counterion displacement is controlled by diffusion, the dielectric response is basically characterized by  the mean relaxation time $\tau_0 \propto L^2/D_{in}$, where $L$ is the associated length scale, and $D_{in}$ is the diffusion constant of counterions which is sufficiently well approximated by the diffusion constant of bulk ions \cite{Colby,Wong}. Since we deal with Na-DNA solutions, we take the diffusion
constant of Na$^+$ ions $D_{in} = 1.33 \cdot 10^{-9}$ m$^2/$s. Note that the
equation $\tau_0 \propto L^2/D_{in}$ is a scaling relationship and the
proportionality constant is of order one (see Discussion, part B).

Several length scales are theoretically expected to be associated with 
dielectric relaxations of polyelectrolytes  in solution: the contour length, 
the Debye screening length, the polymer chain statistical segment length and the 
polymer solution mesh size.

The mean relaxation time for pure water DNA solutions, as well as for DNA 
solutions with added salt, is found in the range $10^{-8} - 1.5\cdot 10^{-6}$~s for the HF mode and $2\cdot 10^{-6} - 3\cdot 10^{-4}$~s for the LF mode. The corresponding characteristic length for the HF mode spans the range from 3 nm to 50 nm, while the characteristic length for the LF mode varies between 50 nm and 750 nm. 
Both of them are thus not within the range of the contour length distribution in our samples.  
On the other hand, the Debye screening length, the polymer solution mesh size and the polymer statistical segment length appear as plausible candidates. Moreover, it is noteworthy that specifically the 
values of the characteristic length for the LF mode are close to the values expected for the DNA persistence length.

%%%%%%%%%%%%%%FIGURE%%%%%%%%%%%%%%%%%%%%%%%%
\begin{figure}
\resizebox{0.48\textwidth}{!}{\includegraphics*{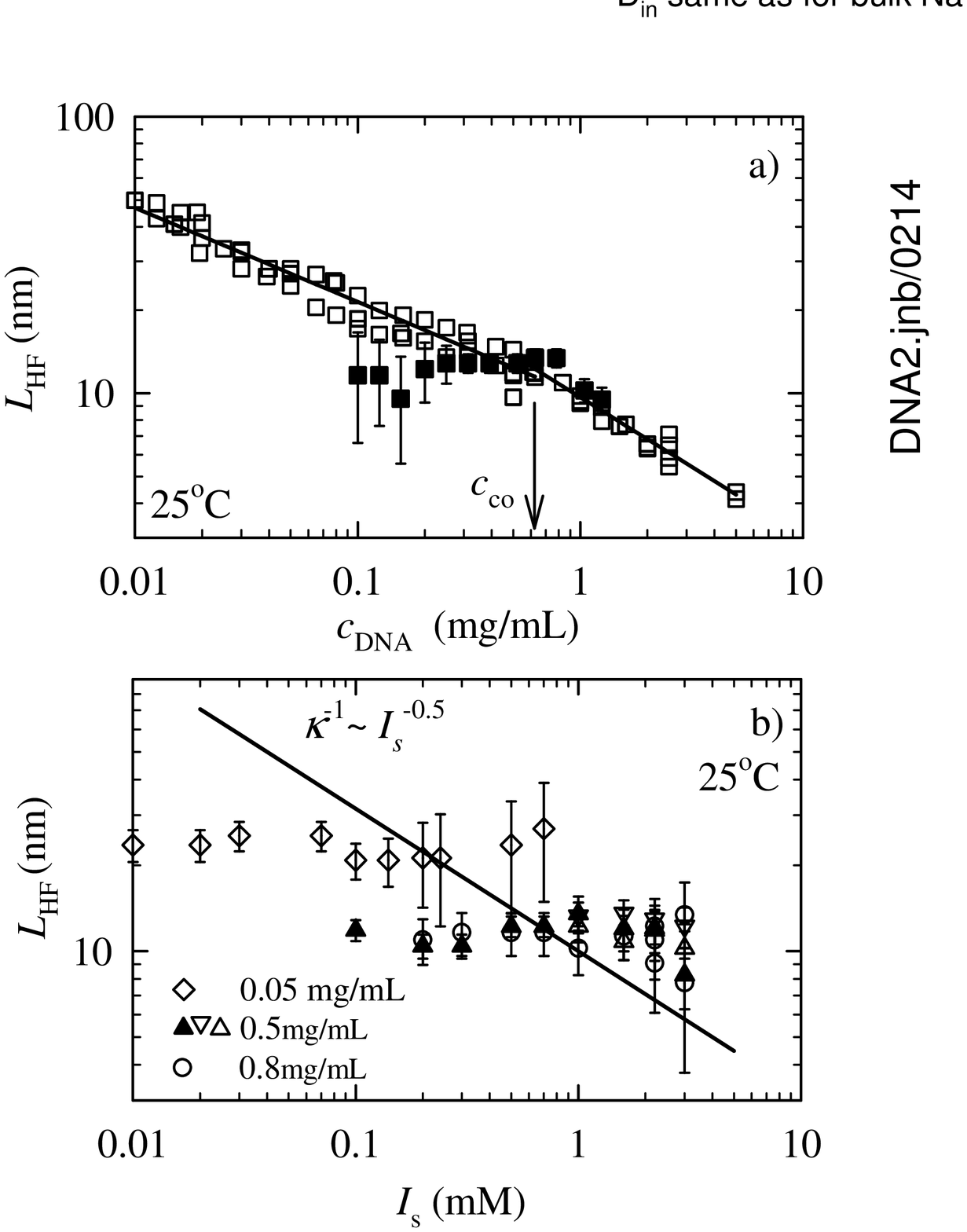}}
%Fig6_0214.JPG
\caption{
a) Characteristic length of the HF mode ($L_{\textrm{HF}}$) for pure water DNA solutions (protocol 
I, open squares) and for DNA solutions with added salt $I_s = 1$~mM (protocol 
II$3$, full squares) as a function of DNA concentration ($c_{\textrm{DNA}}$). The full line is a fit to the power law $L_{\textrm{HF}} \propto$$c_{\textrm{DNA}}^{-0.33}$ and $\propto$$c_{\textrm{DNA}}^{-0.5}$ for  $c_{\textrm{DNA}}$ smaller and larger than $c_{\textrm{co}}\sim 0.6$ mg/mL, respectively. b) Characteristic length of the HF mode ($L_{\textrm{HF}}$) for DNA solutions with varying added salt ($I_s$) for three representative DNA concentrations: $c_{\textrm{DNA}}=0.05$~mg/mL (diamonds, protocol II$1$), $c_{\textrm{DNA}}=0.5$~mg/mL (full triangles, protocol II$1$; open triangles, protocol II$2.1$; open inverse triangles, protocol II$2.2$) and $c_{\textrm{DNA}}=0.8$~mg/mL (circles, protocol II$2.1$). The full line denotes Debye screening length $\kappa^{-1}$ for the investigated range of added salt ionic strength $I_s$.
}
\label{fig6}
\end{figure}
%%%%%%%%%%%%%%FIGURE%%%%%%%%%%%%%%%%%%%%%%%%

\subsection{HF mode}

First, we address the HF mode. In the case of  pure water DNA solutions the characteristic length $L_{\textrm{HF}}$ shows the scaling                              
$L_{\textrm{HF}} \propto c_{\textrm{DNA}}^{-0.5}$ with respect to the DNA concentration, all the way down to a crossover concentration $c_{\textrm{co}}\sim 0.6$ mg/mL. At that point the scaling form is then changed to $L_{\textrm{HF}} \propto c_{\textrm{DNA}}^{-0.33}$ (Fig.~\ref{fig6}a). The observed behavior at high DNA concetrations conforms exactly to the de Gennes-Pfeuty-Dobrynin (dGPD) \cite{Dobrynin,deGennes76,Pfeuty78}  scaling form valid for salt-free polyelectrolyte solutions \cite{Dobrynin,Colby}. At low DNA concetrations, in the regime below $c_{\textrm{co}}$,  $L_{\textrm{HF}}$ displays an unusual behavior generally not observed in semidilute solutions. It appears as though at low DNA concentrations local conformational fluctuations partially expose the hydrophobic core of DNA so that the correlation length scales as $c_{\textrm{DNA}}^{-0.33}$, as is the case of charged chains with partially exposed hydrophobic cores \cite{Dobrynin}. In dilute solutions, but only in dilute solutions, this scaling form would be typical for the average separation between chains \cite{Dobrynin}. The observed crossover might be thought to reflect the border between dilute and semidilute solutions \cite{Ito,Colby} corresponding to the crossover concentration $c^\ast$ \cite{deGennes76}. For the shortest fragments of 2 kbp in our DNA solutions, we get $c^\ast$ of the order of 0.006 mg/mL, while the lowest concentration of DNA solutions is 0.01 mg/mL. In this manner, the interpretation of $c_{\textrm{co}}\sim 0.6$ mg/mL as the dilute-semidilute crossover concentration $c^\ast$ is ruled out.

  With added 1 mM  salt, the dGPD behavior of $L_{\textrm{HF}}$ remains unchanged, thus $L_{\textrm{HF}} \propto c_{\textrm{DNA}}^{-0.5}$, as long as the concentration of intrinsic counterions $c_{in}$ (proportional to $c_{\textrm{DNA}}$) is larger than the concentration of  added salt ions $2I_s$ (Fig.~\ref{fig6}a). When the concentration of intrinsic counterions becomes smaller than the added salt concentrations, the $L_{\textrm{HF}}$  apparently shows a leveling off , with a limiting value close to the Debye length appropriate  for this salt concentration. One should be cautious here since the data become much less reliable exactly at low DNA concentrations (see error bars in Fig.~\ref{fig6}a). Three sets of additional data (Fig.~\ref{fig6}b) for  three representative DNA concentrations with varying added salt also seem to reveal that $L_{\textrm{HF}}$ does not vary with $I_s$ in most of the  {\em measured} range of added salt, while the corresponding Debye screening length
in the  same range of added salt values decreases substantially.  Unfortunately the accuracy of the data again becomes much less reliable due to the progressive merging of the HF  and LF modes when one approaches the regime,  where the  characteristic length scale becomes apparently larger than the  nominal Debye length at that salt concentration.

%%%%%%%%%%%%%%FIGURE%%%%%%%%%%%%%%%%%%%%%%%%
\begin{figure}
\resizebox{0.48\textwidth}{!}{\includegraphics*{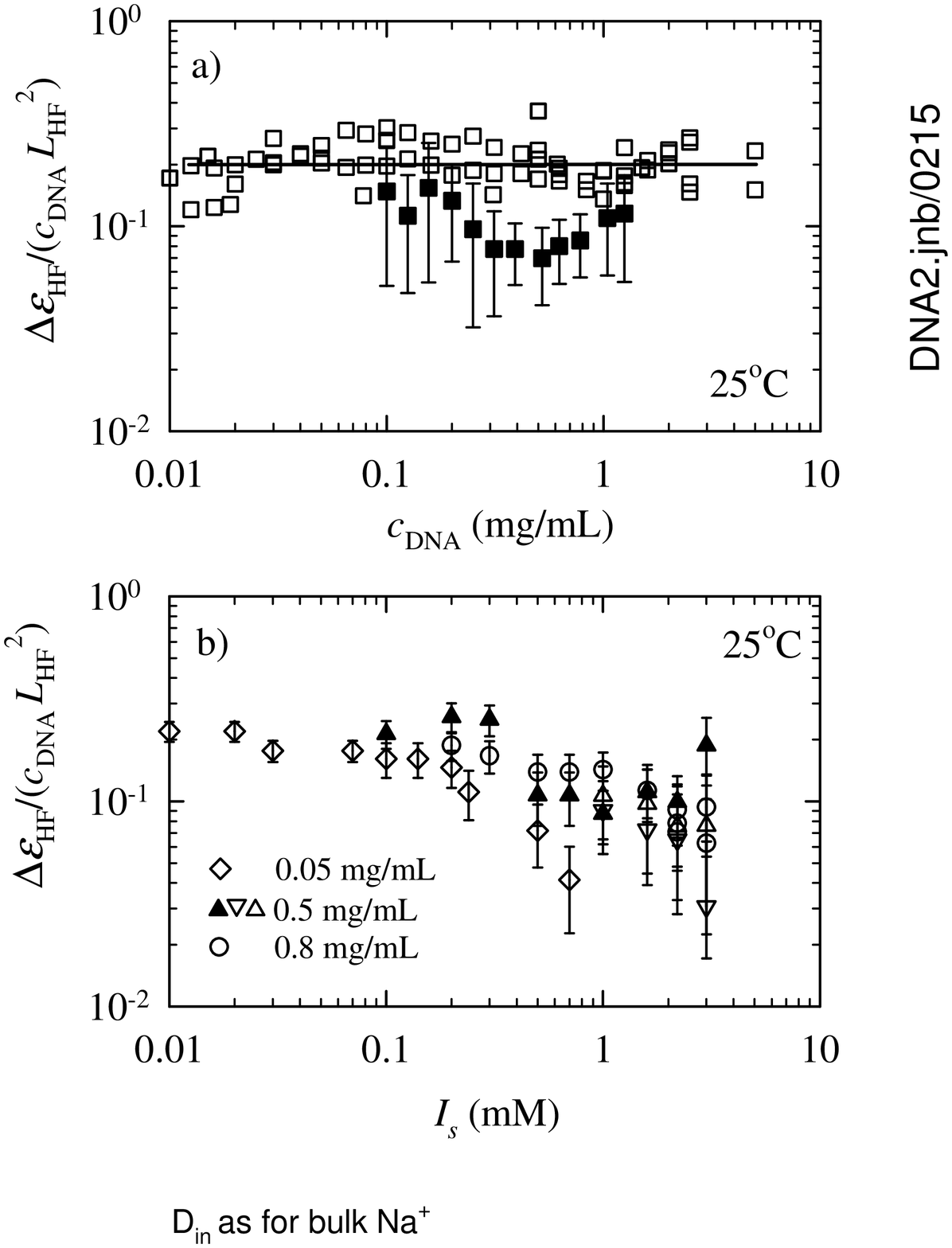}}
%Fig7_0215.JPG
\caption{
a) Normalized dielectric strength of the HF 
mode $\Delta\varepsilon_{\textrm{HF}}/(c_{\textrm{DNA}}\cdot L_{\textrm{HF}}^2)$ as a function of DNA concentration ($c_{\textrm{DNA}}$) for pure water DNA solutions 
(protocol I, open squares) and for DNA solutions with added salt $I_s = 1$~mM 
(protocol II$3$, full squares). The full line is a guide for the eye. b) Normalized dielectric strength of the HF mode $\Delta\varepsilon_{\textrm{HF}}/(c_{\textrm{DNA}}\cdot L_{\textrm{HF}}^2)$  \vs{} ionic strength of the added salt ($I_s$) for three representative DNA concentrations: $c_{\textrm{DNA}}=0.05$~mg/mL (diamonds, protocol II$1$), $c_{\textrm{DNA}}=0.5$~mg/mL (full triangles, protocol II$1$; open triangles, protocol II$2.1$; open inverse triangles, protocol II$2.2$) and $c_{\textrm{DNA}}=0.8$~mg/mL (circles, protocol II$2.1$). }
\label{fig7}
\end{figure}
%%%%%%%%%%%%%%FIGURE%%%%%%%%%%%%%%%%%%%%%%%%

Next we consider the behavior of dielectric strength defined as
$\Delta\varepsilon_{\textrm{HF}} \approx f_{\textrm{HF}} \cdot c_{in} \cdot 
\alpha_{\textrm{HF}}$, where $f_{\textrm{HF}}$ is the fraction of counterions 
participating in the HF process, $c_{in} \left[\textrm{mM}\right] = c_{\textrm{DNA}} \left[\textrm{mg/mL}\right]   \times$ 3 $\mu$mol/mg (as explained in Section~\ref{sec2}) and 
$\alpha_{\textrm{HF}}$ is the corresponding polarizability. The polarizability 
$\alpha_{\textrm{HF}}$ is given by the scaling form $\alpha_{\textrm{HF}} 
\propto e^2 \cdot L_{\textrm{HF}}^2 \cdot / (\varepsilon_0 kT) \propto l_B \cdot 
\varepsilon  \cdot L_{\textrm{HF}}^2$, \cite{Ito,Colby}. Therefore, 
the fraction of counterions $f_{\textrm{HF}}$ participating in the HF process is proportional to $\Delta\varepsilon_{\textrm{HF}}/(c_{\textrm{DNA}}\cdot L_{\textrm{HF}}^2)$. In 
Fig.~\ref{fig7} we show the dependence of  $f_{\textrm{HF}}$ on DNA 
concentration (panel a)) and on the ionic strength of added salt ions (panel 
b)).   	

The $f_{\textrm{HF}}$ data for pure water DNA solution displayed in 
panel a) indicate that the fraction of counterions participating in this relaxation process 
does not depend on the concentration of DNA. Since the HF relaxation happens at the 
length scale $\xi$  which describes the density correlations between DNA chains, this result  indicates that it is the free counterions as opposed to condensed counterions, that can hop from chain to chain in the volume $\xi^3$, that are the relaxation entities 
participating in the HF process. It is noteworthy that a similar interpretation 
was previously proposed by Ito \al{} \cite{Ito} for the relaxation in synthetic 
polyelectrolytes observed in the same frequency range \cite{Colby}. 

The data displayed in Fig.\ref{fig7} b) suggest that the fraction of intrinsic counterions 
$f_{\textrm{HF}}$ active in the HF mode remains constant when salt is added to 
the DNA solution as long as $c_{\textrm{DNA}}$ is substantially larger than 
$I_s$.  However, as soon as the concentration of added salt ions prevails over 
the concentration of intrinsic counterions, $f_{\textrm{HF}}$ starts to 
decrease. This is also discernible in the behavior of the $f_{\textrm{HF}}$ data 
for $I_s = 1$~mM added salt solution shown in panel a). A plausible suggestion 
would be that the salt renormalization of  $f_{\textrm{HF}}$ is a consequence of 
screening due to added salt ions that seem to diminish the effective number of 
counterions that can participate in the chain - chain hopping process.

\subsection{LF mode}

\begin{figure}
\resizebox{0.48\textwidth}{!}{\includegraphics*{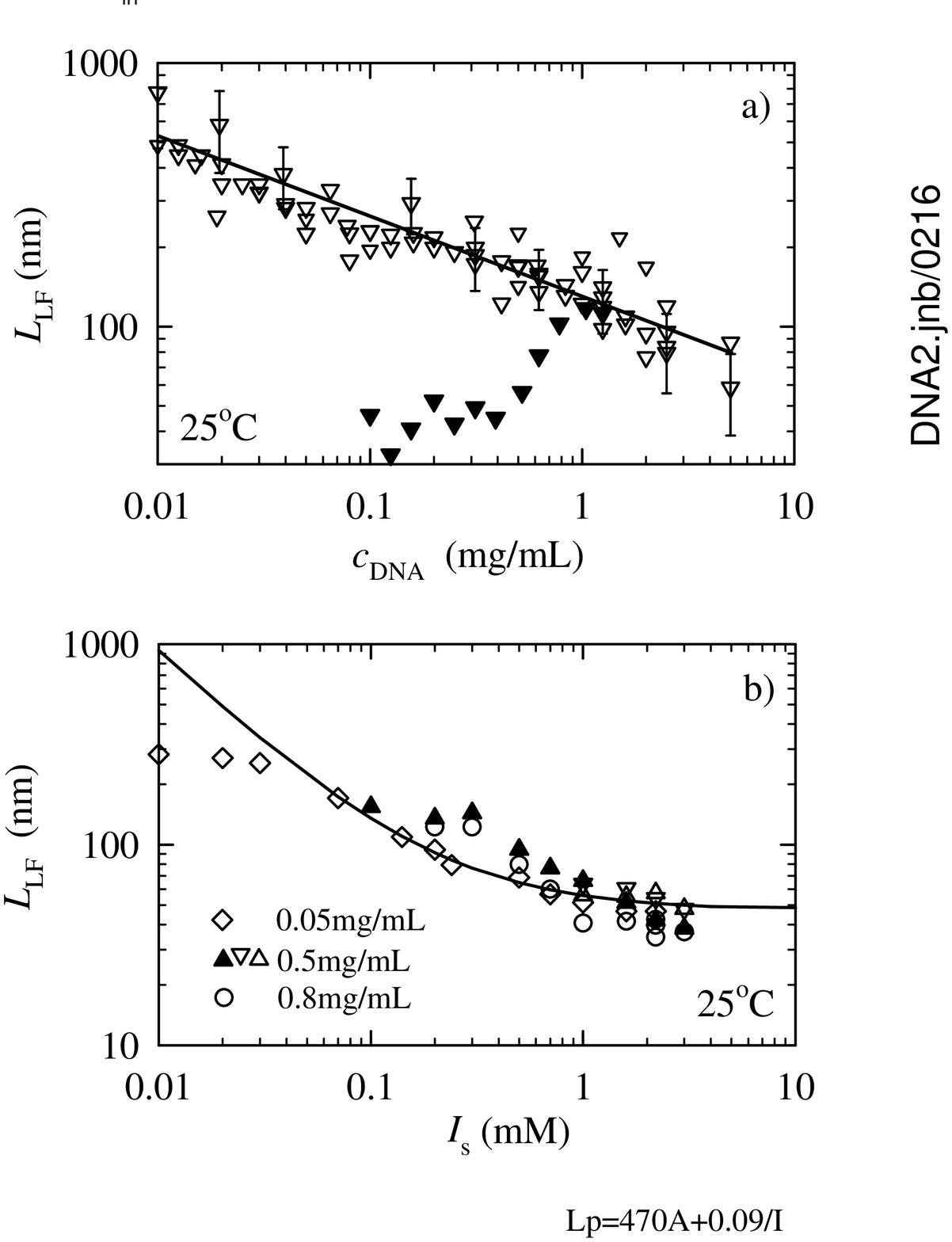}}
%Fig8_0216.JPG
\caption{
a) Characteristic length of the LF mode ($L_{\textrm{LF}}$) for pure water DNA solutions (protocol I, open inverse triangles) and for DNA solutions with added salt $I_s = 1$ mM (protocol II$3$, full inverse triangles) as a function of DNA concentration ($c_{\textrm{DNA}}$). The full line is a fit to the power law $L_{\textrm{LF}} \propto c_{\textrm{DNA}}^{-0.29\pm0.04}$. b) Characteristic length of the LF mode ($L_{\textrm{LF}}$) for DNA solutions with varying added salt ($I_s$) for three representative DNA concentrations: $c_{\textrm{DNA}}=0.05$~mg/mL (diamonds, protocol II$1$), $c_{\textrm{DNA}}=0.5$~mg/mL (full triangles, protocol II$1$; open triangles, protocol II$2.1$; open inverse triangles, protocol II$2.2$) and $c_{\textrm{DNA}}=0.8$~mg/mL (circles, protocol II$2.1$). The full line is a fit to the expression $L_p = L_0 + a \cdot I_s^{-1}$ with $L_0 = 470$~\AA~ and $a =  0.09$~\AA.
}
\label{fig8}
\end{figure}

Second, we address the LF mode. For 
pure water DNA solutions (protocol I), the characteristic length 
$L_{\textrm{LF}}$ increases with decreasing DNA concentration in almost three 
decades wide concentration range (open inverse triangles in Fig.~\ref{fig8}a)) 
following the power law $L_{\textrm{LF}} \propto c_{\textrm{DNA}}^{-0.29\pm0.04}$. The exponent $-0.29\pm0.04$ suggests that  in this regime $L_{\textrm{LF}}$ is proportional to the average size of the polyelectrolyte chain that behaves as  a random walk of correlation blobs and scales as $ c_{\textrm{DNA}}^{-0.25 }$ \cite{Dobrynin}. For DNA solutions with added salt $I_s = 1$ mM (protocol II$3$, full inverse triangles in Fig.~\ref{fig8}a)), 
$L_{\textrm{LF}}$ coincides with the one found for pure water solutions with 
high DNA concentrations. As soon as the concentration of intrinsic counterions 
$c_{in}$ (proportional to $c_{\textrm{DNA}}$) becomes smaller than the 
concentration of bulk ions from added salt $2I_s$, $L_{\textrm{LF}}$ starts to 
deviate from the $L_{\textrm{LF}} \propto c_{\textrm{DNA}}^{-0.29}$ behavior and 
decreases to attain value of about 500 \AA~at which it saturates. 

The dependence of $L_{\textrm{LF}}$ on the added salt ionic strength $I_s$ is 
shown in Fig.~\ref{fig8}b) for three DNA concentrations. The observed data can be 
nicely fit to the OSF behavior \cite{Odijk77,Skolnick77} of the form: 
$L_p = L_0 + a \cdot I_s^{-1}$. We get $L_0 = 470$ \AA~ for the structural persistence length and $a = 0.09$ \AA M. While the value of $L_0$ close to 500 \AA~is in accordance with standard expectations for DNA \cite{Bloomfield00}, the value of the coefficient $a$ is somewhat smaller than expected by the OSF theory $a = 0.324$ \AA M. It is noteworthy that the OSF model applies as long as the ionic strength of added salt is larger than the concentration of the intrinsic counterions. The data for $c_{\textrm{DNA}} = 0.05$, 0.5 and 0.8 mg/mL deviate from the OSF behavior for $I_s < 0.03$~mM,  $I_s < 0.3$ mM and $I_s < 0.5$ mM, respectively. The value of $L_{\textrm{LF}}$ in this low salt limit attains the same value as in pure water DNA solutions (see Fig.~\ref{fig8}a).

%%%%%%%%%%%%%%FIGURE%%%%%%%%%%%%%%%%%%%%%%%%
%%%%%%%%%%%%%%FIGURE%%%%%%%%%%%%%%%%%%%%%%%%

%%%%%%%%%%%%%%FIGURE%%%%%%%%%%%%%%%%%%%%%%%%
\begin{figure}
\resizebox{0.48\textwidth}{!}{\includegraphics*{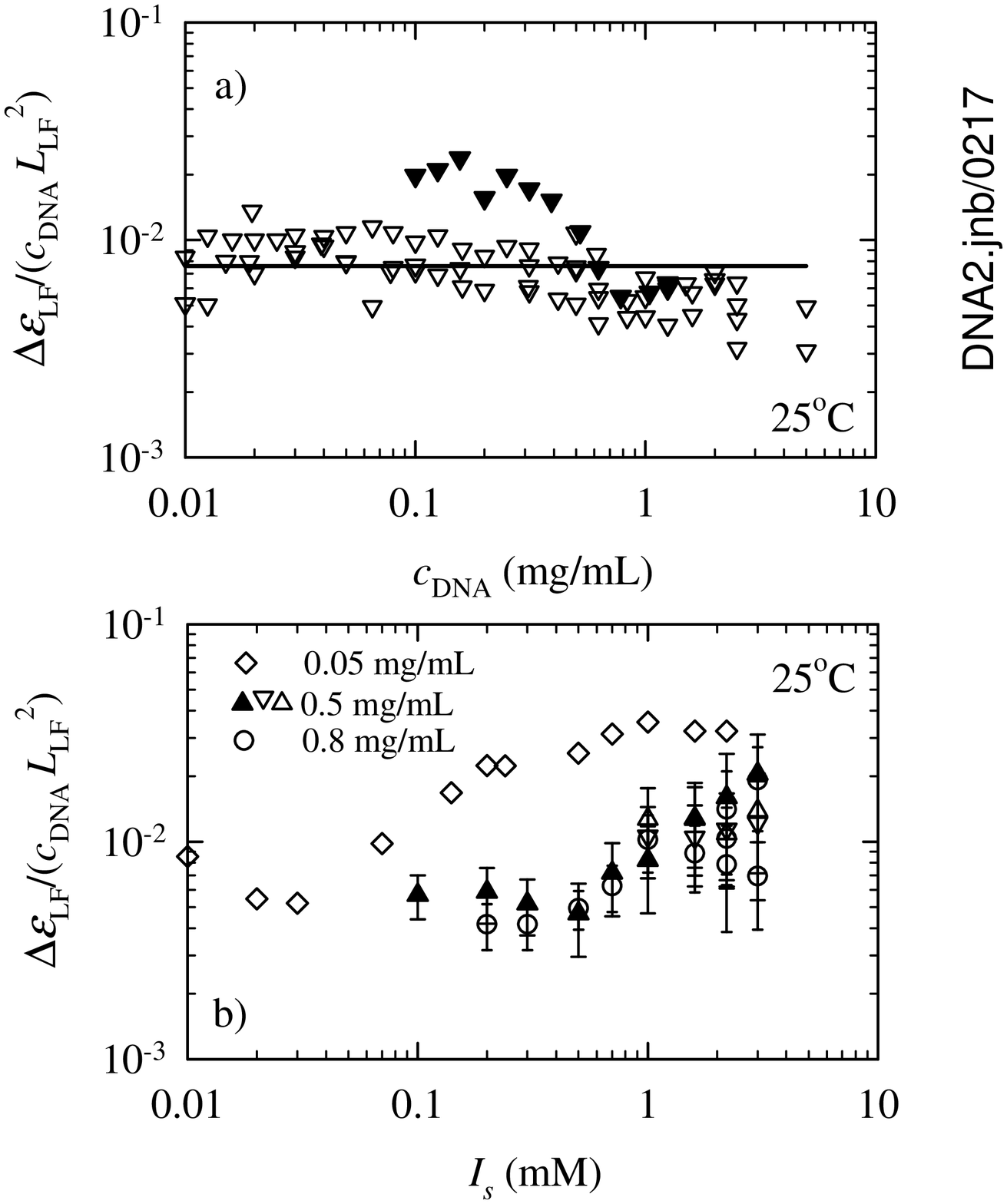}}
%Fig9_0217.JPG
\caption{
a) Normalized dielectric strength of the LF mode 
$\Delta\varepsilon_{\textrm{LF}}/(c_{\textrm{DNA}}\cdot L_{\textrm{LF}}^2)$  as 
a function of DNA concentration ($c_{\textrm{DNA}}$) for pure water DNA 
solutions (protocol I, open inverse triangles) and for DNA solutions with added 
salt $I_s = 1$ mM (protocol II$3$, full inverse triangles).  The full line is a 
guide for the eye. b) Normalized dielectric strength of the LF mode 
$\Delta\varepsilon_{\textrm{LF}}/(c_{\textrm{DNA}}\cdot L_{\textrm{LF}}^2)$  
\vs{} ionic strength of the added salt ($I_s$) for three representative DNA concentrations: $c_{\textrm{DNA}}=0.05$~mg/mL (diamonds, protocol II$1$), $c_{\textrm{DNA}}=0.5$~mg/mL (full triangles, protocol II$1$; open triangles, protocol II$2.1$; open inverse triangles, protocol II$2.2$) and $c_{\textrm{DNA}}=0.8$~mg/mL (circles, protocol II$2.1$).}
\label{fig9}
\end{figure}
%%%%%%%%%%%%%%FIGURE%%%%%%%%%%%%%%%%%%%%%%%%

In Fig.~\ref{fig9}a) we show dependence of  $f_{\textrm{LF}} \propto \Delta\varepsilon_{\textrm{LF}}/(c_{\textrm{DNA}}\cdot L_{\textrm{LF}}^2)$  on the DNA concentration assuming again that the polarizability $\alpha_{\textrm{LF}}$ varies as the characteristic length squared. The data show that the fraction of counterions $f_{\textrm{LF}}$ active in the LF mode is roughly speaking independent of the DNA concentration. This is valid for the pure water DNA solution, as well as in the case of  $I_s = 1$~mM added salt solution. In our view this result together with the fact that the LF relaxation happens at the length scale of the average size of the polyelectrolyte chain suggests that the LF relaxation engages mostly condensed counterions along and in close vicinity of the chain. 

However, the data displayed in panel b), as well as the data for $I_s = 1$~mM in panel a) suggest  that the fraction of  counterions participating in the LF process $f_{\textrm{LF}}$ becomes larger in the case of added salt solutions, compared to the pure water case, if the concentration of added salt ions becomes larger than $c_{\textrm{DNA}}$. 
This means that at least some of the free counterions join the relaxation of the condensed counterions along the segments of the same chain. It is thus impossible to completely separate condensed counterions from free counterions in their contribution to the LF relaxation mode. This conclusion bears crucially on the assumption that the scale of polarizability is given by $L_{\textrm{LF}}(I_s)$. It is noteworthy that for the HF and LF modes the addition of salt changes the effective number of participating counterions in the opposite way. This might be attributed to the increased screening for the interchain HF relaxation, and to the intrinsic counterion atmospheres squeezed closer to the chains due to reduced Debye length for the LF relaxation along the chain. 

As a final remark, we point out that our results showing that the HF relaxation can be attributed to the semidilute mesh size $\xi$, in the polyelectrolyte semidilute solution in which single chain persistence length associated with the LF relaxation is always larger than $\xi$, confirm the prediction of Odijk \cite{Odijk79} that the same scaling law $\xi \propto c^{-0.5}$ should also be valid for semiflexible DNA polymers.

\section{Discussion}

First, let us summarize the results of dielectric spectroscopy measurements. In 
the linear ac field regime, two broad relaxation modes are observed 
corresponding to three different time and length scales. The HF mode is centered 
in the frequency range between 0.1 MHz and 15 MHz, depending solely on the DNA 
concentration as long as the DNA concentration remains larger than the added salt concentration. In this regime, the characteristic length scale is identified with the mesh size, and varies as $\xi \propto c_{\textrm{DNA}}^{-0.5}$.  Our data also seem to indicate that once the added salt becomes
larger than the concentration of DNA intrinsic counterions, the high
frequency characteristic length, rather than scaling as the
semidilute solution correlation length, levels off at a value close
to the corresponding Debye length. More systematic experiments are needed to asses the possible added salt dependence in this regime of salt concentrations. In the limit of low DNA concentrations and low added salt, the semidilute solution correlation length smoothly crosses over to a less rapid scaling as $\propto c^{-0.33}$ probably reflecting the appearance of locally fluctuating regions with exposed hydrophobic cores.

The LF mode is centered in the frequency range between 0.5 kHz$<\nu_{\textrm{LF}}<$70 kHz. In DNA solutions with added salt, the characteristic length scale of this mode 
corresponds to the persistence length, which varies experimentally as $L_p 
\propto I_s^{-1}$. In the limit of low added salt, the characteristic  length scale smoothly merges with the average size of the Gaussian chain composed of correlation blobs, which varies with DNA concentration as $L_{\textrm{LF}}\propto c_{\textrm{DNA}}^{-0.25}$. 

The dielectric data also seem to indicate that the free DNA counterions 
are mostly responsible for the high frequency relaxational mode, whereas the low 
frequency mode appears to be more complicated and the decoupling of MO condensed and free DNA counterions seems to be difficult with any degree of confidence. 

Investigation of dielectric properties of DNA goes back to early '60s and since 
then a reasonable number of papers have been published 
\cite{Nandi00,Colby,Saif91,Mandel77,Sakamoto76,Molinari,Takashima84,Bone95,Lee98}. In 
these studies, two dispersion modes associated with the counterion fluctuations 
were found, one at very low frequencies, which depended on molecular weight 
(\ie{} degree of polymerization, $N$) and another one in the intermediate 
frequency region, which showed no $N$ dependence, but did show a pronounced 
concentration dependence. For the sake of completeness we also mention the third 
relaxation in the GHz frequency range, which is due to solvent (water) 
relaxation and therefore not directly a concern of this paper. 

The low frequency mode was consistently associated with the DNA contour length, 
while the  intermediate frequency one was often associated with the DNA 
statistical segment length. These interpretations were based on the theoretical 
models of the Mandel group \cite{Mandel77,Mandel84,derTouw74} developed for the 
case of a single polyelectrolyte (limit of very dilute solutions).  Another 
interpretation of the intermediate frequency relaxation, proposed and verified 
until now only for synthetic monodisperse polyelectrolytes, was that it was due 
to the counterion fluctuation along the correlation length 
\cite{Colby,Ito,Bordy02}. It is worth noting the suggestion of Odijk 
\cite{Odijk79} that the mesh size gives a physical meaning to the statistically 
independent chain segment length, which also scales as $ c^{-0.5}$ 
\cite{Mandel77}. The segment length was postulated to be due to the potential 
barriers along the chain, whose origin might be in the correlation length that 
measures the mean distance between contact points of the overlaping chains \cite 
{deGennes76}.

Our work reveals the existence of {\bf two}, rather than one, distinct relaxation modes in the {\em s.c.} intermediate frequency range due to either a correlated response of counterions in the mesh of DNA chains in the solution, what we call the HF mode, or due to a response of the counterions along a single DNA chain, what we refer to as the LF mode.

No previous experimental work was able to distinguish 
these two concentration dependent dispersions. The reason probably lies in the fact that the data analysis performed in the reported DNA dielectric spectroscopy studies 
was not powerful enough to reveal and characterize two modes so close in frequency, 
where in addition one of them is small in amplitude. Indeed, work by Lee and 
Bone \cite{Lee98} hinted at two overlapping dispersions, but the authors were 
not able to characterize the smaller one properly.  Another reason lies in the 
fact that none of these investigations covered so wide a range of DNA 
concentrations and added salt ionic strengths, as we did in this work.

\subsection{Conformation of DNA in low salt solutions}

In this work we have paid special attention to the issue of the stability of 
ds-DNA helix, \ie{} to the denaturation phase diagram \cite{Bloomfield00}. The 
issue of DNA conformation in pure water solutions is of paramount importance for 
proper understanding of our experiments. The question here is whether DNA at 
very low salt conditions is in the double stranded or single stranded form. 
Dielectric spectroscopy results strongly indicate that the double stranded form 
of DNA is stable in all pure water \cite{purewater} solutions studied. Of course, a precise and definitive information on the polyelectrolyte intrachain conformation of DNA solutions would demand the small-angle neutron scattering and/or X-ray scattering experiments performed at the same conditions. 

First evidence for the stability of the ds-DNA form comes from the fact that the 
results of the dielectric spectroscopy measurements obtained in DNA solutions 
with added salt prepared from water (protocol II$1$) coincide with the results 
obtained on DNA solutions prepared from 10 mM NaCl (protocols II$2$, II$3$) (see 
Fig.~\ref{fig6}b), \ref{fig7}b), \ref{fig8}b) and \ref{fig9}b)). Second, we mention that Mandel \cite{Mandel77} already reported that dielectric behavior in pure water DNA solutions was not found to differ markedly from that at low salt concentrations. Also the measured osmotic coefficient in nominally pure water conditions \cite{Auer69} confirms 
the assumption of an intact double stranded DNA form. Since the Manning charge 
density parameters for ss- and ds-DNA are so different this difference should be 
apparent also in the measured osmotic coefficient.  None is detected however. 

\begin{figure}
\resizebox{0.48\textwidth}{!}{\includegraphics*{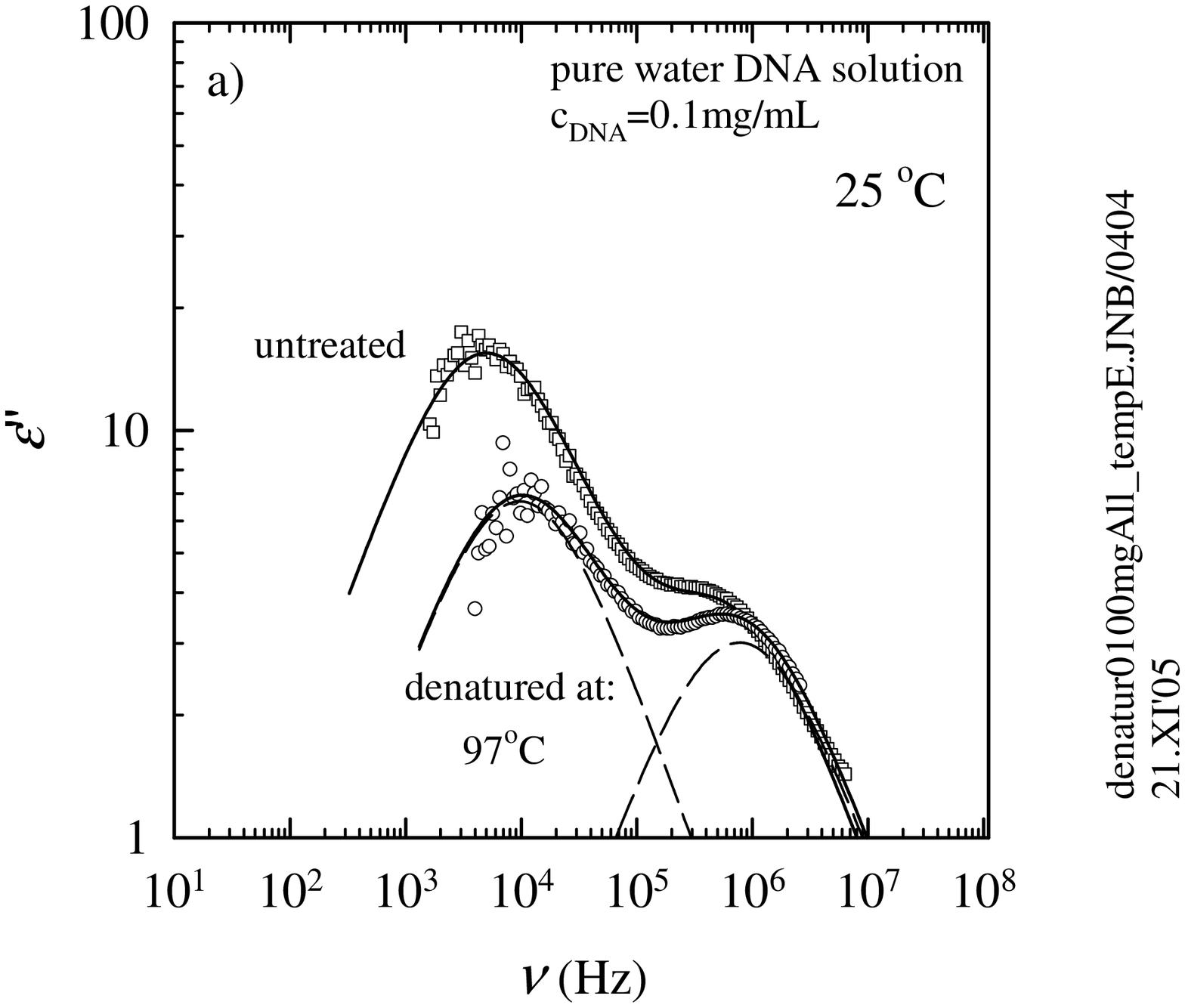}}
%denaturationE0404
\resizebox{0.48\textwidth}{!}{\includegraphics*{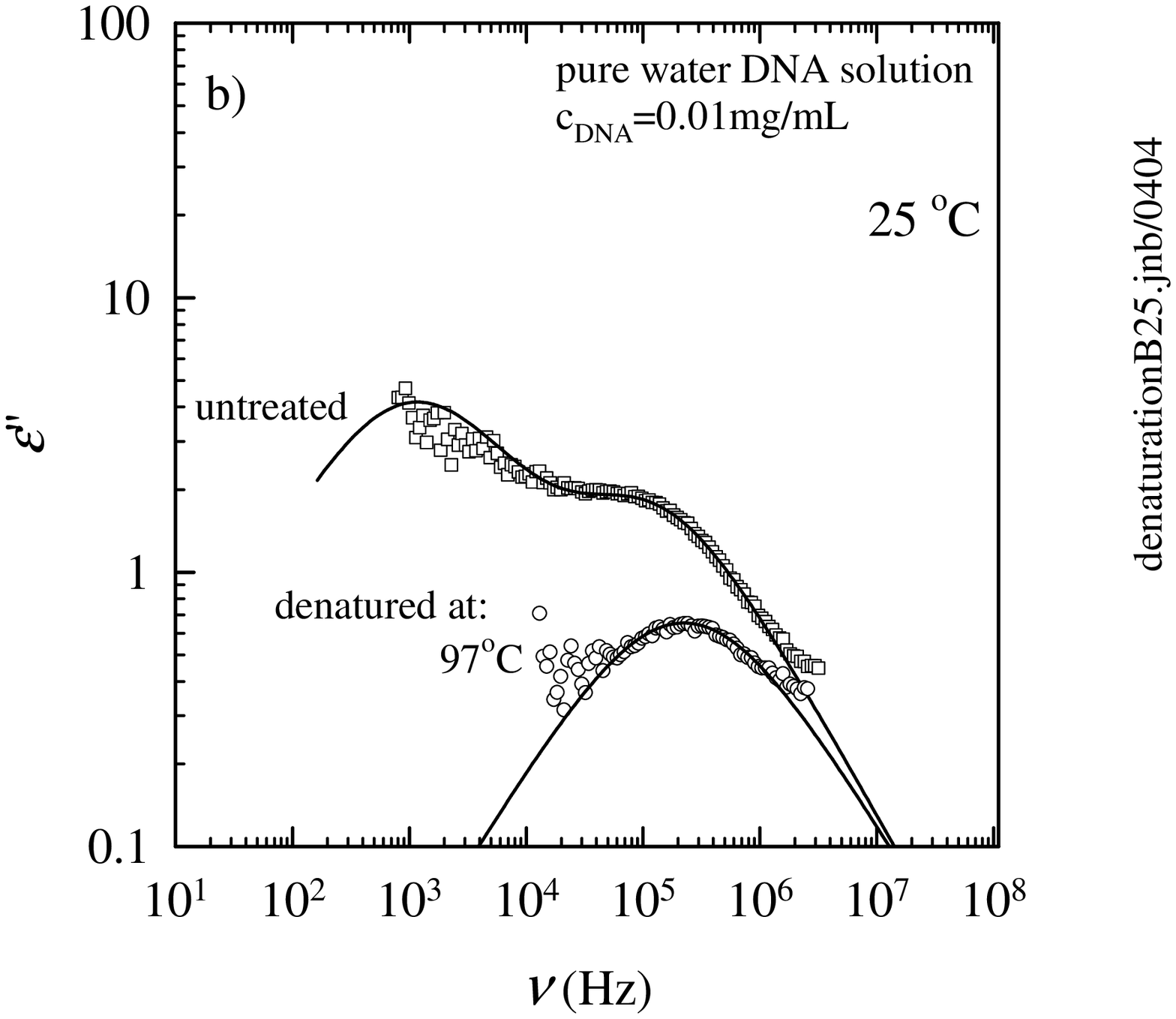}}
%denaturationB0404
\caption{ Double logarithmic plot of the frequency dependence of the imaginary part of the 
dielectric function ($\varepsilon$'') at T=25$^{o}$C of a pure water DNA solution (protocol I) with DNA concentration $c_{\textrm{DNA}}=0.1$ mg/mL (panel a)) and $c_{\textrm{DNA}}=0.01$ mg/mL (panel b)) before (denoted as untreated) and after the heating to 97$^{o}$C (denoted as denaturated, see Text). The full lines are fits to the sum of the two HN forms; the dashed lines represent a single HN form.}
\label{fig10}
\end{figure}

In an attempt to clarify more this issue, we have measured dielectric properties 
of pure water DNA solutions for the DNA concentration range between 
$c_{\textrm{DNA}}=0.5$ mg/mL and $c_{\textrm{DNA}}=0.01$ mg/mL, prepared according to protocol I, before and after the controlled denaturation protocol. Denaturation was accomplished by the heating of solutions for 20 min at temperature of 97$^{o}$C, 
followed by quenching to 4$^{o}$C. Dielectric measurements were subsequently made at 
25$^{o}$C. The observed results were similar for all studied DNA solutions. The 
dielectric strength and the relaxation time of the LF mode decreased 
substantially after the heating, while the change observed for the HF mode was 
much smaller (Fig.~\ref{fig10}).  For $c_{\textrm{DNA}}=0.01$ mg/mL, the LF mode was not observed at all after the denaturation, implying the dielectric strength $\Delta\varepsilon<1$ (see Fig.~\ref{fig10}b)). Notice that $\Delta\varepsilon$ of the LF 
mode for this DNA concentration is about 10.

%%%%%%%%%%%%%%FIGURE%%%%%%%%%%%%%%%%%%%%%%%%
%%%%%%%%%%%%%%FIGURE%%%%%%%%%%%%%%%%%%%%%%%%

%%%%%%%%%%%%%%FIGURE%%%%%%%%%%%%%%%%%%%%%%%%

\begin{figure}
\resizebox{0.47\textwidth}{!}{\includegraphics*{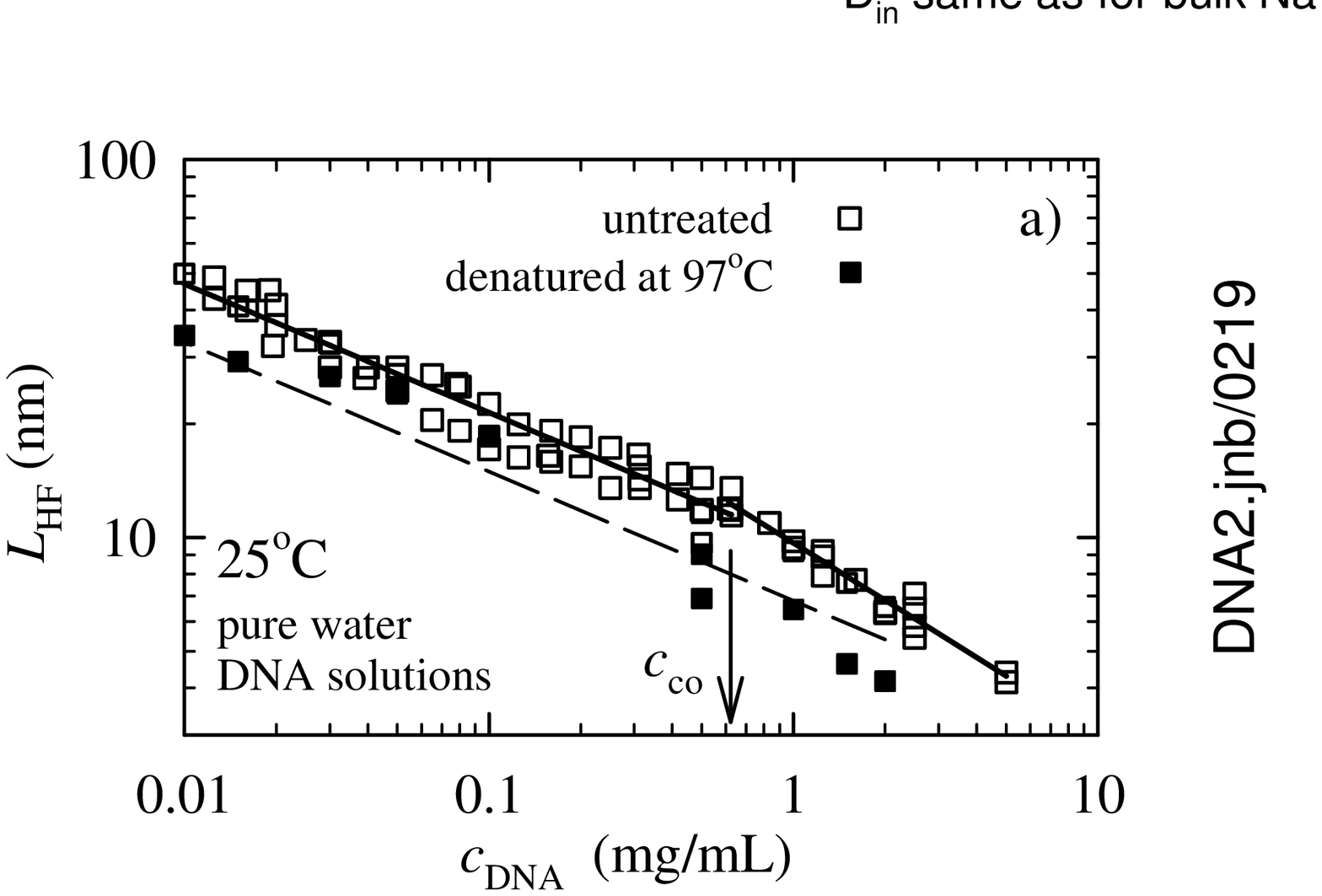}}
\resizebox{0.48\textwidth}{!}{\includegraphics*{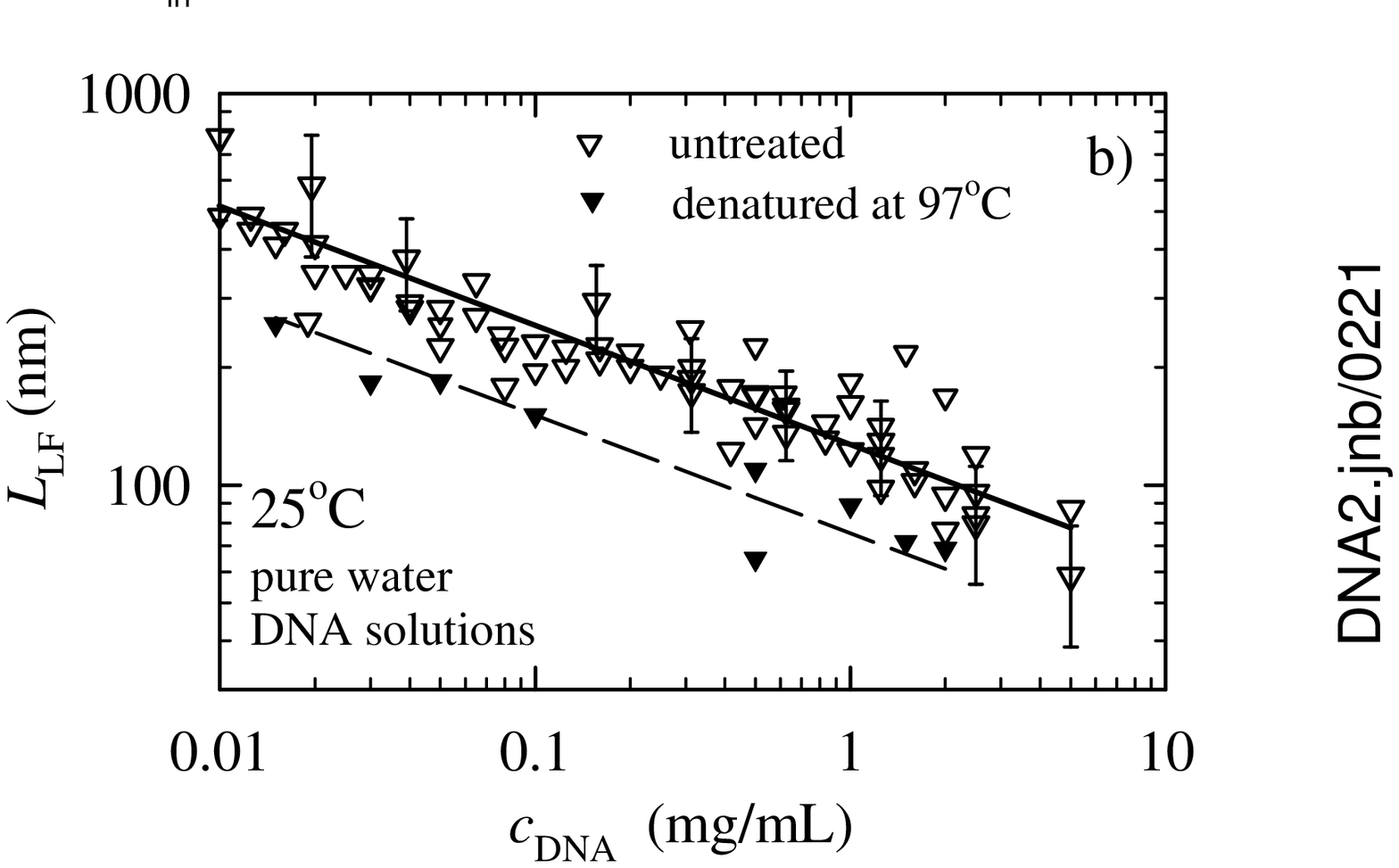}}
%0219_0221.jpg
\caption{ a) The characteristic length $L_{\textrm{HF}}$ of the HF mode before (open squares) and after the heating to 97$^{o}$C (full squares). The full line is a fit to the power law $L_{\textrm{HF}} \propto$$c_{\textrm{DNA}}^{-0.33}$ and $\propto$$c_{\textrm{DNA}}^{-0.5}$ for  $c_{\textrm{DNA}}$ smaller and larger than $c_{\textrm{co}}\sim 0.6$ mg/mL, respectively. The dashed line is a fit to the power law $L_{\textrm{HF}} \propto$$c_{\textrm{DNA}}^{-0.33}$ in the whole DNA concentration range. b) The characteristic length  $L_{\textrm{LF}}$ of the LF mode before (open triangles) and after the heating to 97$^{o}$C (full triangles). The full line and dashed lines are fits to the power law $L_{\textrm{LF}} \propto c_{\textrm{DNA}}^{-0.29\pm0.04}$.}
\label{figNEW1}
\end{figure}
%%%%%%%%%%%%%%FIGURE%%%%%%%%%%%%%%%%%%%%%%%%

We note that $L_{\textrm{LF}}$, which measures the average size of the DNA chain,  decreased after the heating, but showed the same power law behavior $L_{\textrm{LF}} \propto c_{\textrm{DNA}}^{-0.29\pm0.04}$ as for the untreated DNA solution (Fig.~\ref{figNEW1}b)). The observed change indicates that denatured ss-DNA, which is in the form of coil, is shorter and has smaller average size of the chain. Furthermore, $L_{\textrm{HF}}$  which measures the correlation length of the DNA solution, decreased after the heating and showed the power law behavior $L_{\textrm{HF}} \propto$$c_{\textrm{DNA}}^{-0.33}$ in the whole DNA concentration range (Fig.~\ref{figNEW1}a)). A smaller value of $L_{\textrm{HF}}$ is expected for the denatured DNA solution, since such a solution should contain twice the number chains. The power law behavior with the exponent -0.33, observed now also for the larger DNA concentrations, indicates that the hydrophobic core of DNA is fully exposed once the heating protocol was applied. 

All these results confirm that although untreated DNA at low salt and semidilute conditions might show locally exposed hydrophobic cores in a  dynamic sense, a real/complete unzipping and separation of the strands might be accomplished only after the denaturation/heating protocol is applied. These observations therefore  suggest that DNA in pure water solutions is not denatured into two spatially well separated single strands, but is rather in the double stranded form, locally interspersed with exposed hydrophobic cores in the limit of low DNA concentrations.

Furthermore, even after denaturation the two DNA strands appear to remain in relatively close proximity rather than becoming completely dissociated, an observation well substantiated also by the correlation length measurements by SANS at semidilute DNA conditions \cite{Hammouda06A}. In these measurements the correlation length measures the characteristic distance between the hydrogen-containing (sugar-amine base) groups. Hammouda and Worcester \cite{Hammouda06A}  have recently determined that on melting of DNA in DNA/d-ethylene-glycol mixtures the correlation length increases from about 8 \AA~ to about 12-15\AA, which implies that  
even after melting the two strands of the ds-DNA remain in very close proximity. This appears to be due to the presence of other chains in the semidilute solution that spatially constrain the separate strands even after melting and prevent complete dissocation \cite{Hammouda06B}.  

The question furthermore arises if the renaturation process is fast 
enough to occur partially during dielectric measurements which are performed at 
25$^{o}$C. In order to check this, we have repeatedly measured the response of 
pure water DNA solution, after the heating protocol to 97 $^{o}$C was applied. 
The time span was 100 min, the first measurement was taken 3 minutes after the 
sample was heated from 4$^{o}$C to 25 $^{o}$C. The observed response after 3 min and after 100 min was the same (inside the error bar of 1.5\%) indicating very long time constant characterizing the renaturation process. This result confirmed 
that the observed dielectric properties are the ones of ss-DNA. 

Osmotic pressure results of Raspaud \al{} \cite{Raspaud00} on short nucleosomal 
fragments of DNA also indicate that at high enough concentration, intrinsic DNA 
counterions prevent destabilization of ds-DNA helix in pure water solutions.  
These experiments show that the role of added salt ($I_s$) in the ds-DNA 
stabilization becomes negligible in the high DNA concentration range \ie{} in 
the limit $c_{\textrm{DNA}} \gg I_s$. Osmotic pressure of intrinsic DNA counterions can 
thus be high enough to prevent denaturation of ds-DNA helix in pure water 
solutions. 

UV spectrophotometry experiments performed previously by Record \cite{Record75} 
on T4 and T7 phage DNA have also shown that ds-DNA denaturation depends not only 
on added salt concentration $I_s$, but also  on the 
concentration of intrinsic DNA counterions, $c_{in} \left[\textrm{mM}\right] = c_{\textrm{DNA}} \left[\textrm{mg/mL}\right]   \times$ 3 $\mu$mol/mg 
(see Section~\ref{sec2}). Moreover, ds-DNA was found to be stable at 25$^{o}$C 
dissolved in nominally pure water (no added salt, $I_s \rightarrow 0$) for the 
concentration of intrinsic counterions larger than 0.2 mM. This again is 
consistent with our experiments on dielectric relaxation of DNA 
solutions. 

Another meaningful question would be, what is the smallest $c_{in}$ which can still keep
DNA in the double stranded form. The UV absorbance results indicated that no
added salt is needed to stabilize ds-DNA in the case when the intrinsic
counterion concentration, $c_{in}$, is larger than 0.2 mM, while dielectric
spectroscopy results suggest it can be one order of magnitude smaller,
$c_{in} > 0.03$~mM. Notice that in the latter case $c_{in}$ is still larger than 
the estimated ion concentration of pure water ($2I_s=0.02$ mM), \ie{} $c_{in} > 
2I_s$. The osmotic pressure data indicate \cite{Raspaud00}  that intrinsic 
counterions themselves can stabilize ds-DNA for $c_{in}$ larger than 30 mM and 
150 mM for two different added salt concentrations $2I_s = 4$ mM and 20 mM, 
respectively. All these results suggest that the concentrations of intrinsic 
counterions and added salt themselves are less important, rather it is their 
ratio that defines which limit prevails. Thus absorbance and osmotic pressure 
data both suggest that for $c_{in}/2I_s >10$ intrinsic counterions prevail in 
stabilizing ds-DNA helix, while dielectric results suggest that this ratio might 
be shifted to 
$c_{in}/2I_s >1$. Nevertheless, the unusual scaling exponent -0.33 of the 
semidilute correlation length found for the DNA pure water solutions with DNA 
concentrations smaller than $c_{\textrm{co}}\sim 0.6$ mg/mL seem to suggest the 
existence of local conformational fluctuations which partially expose the 
hydrophobic core of DNA. In order to verify this proposal a further comparative 
study, including UV spectrophotometry and dielectric spectroscopy, is planned to 
study in depth the conformation of DNA in pure water solutions.  We are however 
convinced that in all our investigations DNA did not exist as two separated 
single coils, rather DNA was effectively always in its double stranded form.

\subsection{Ionic screening: added salt {\em versus} intrinsic DNA counterions}
	 
\begin{figure}
\resizebox{0.48\textwidth}{!}{\includegraphics*{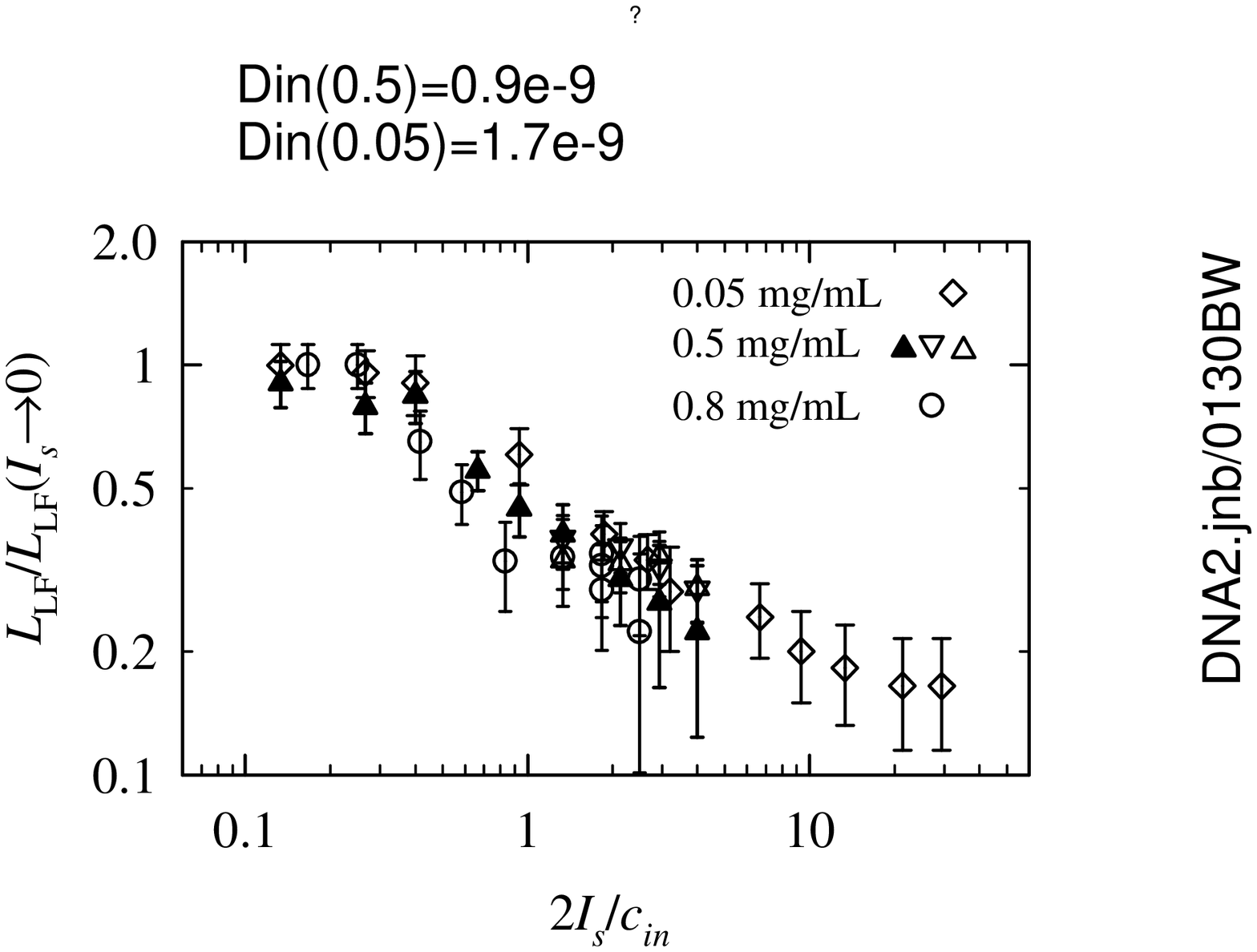}}
%0130.eps
\caption{
Characteristic length of the LF mode ($L_{\textrm{LF}}$) normalized with the value in pure 
water solutions (low salt limit $I_s = 0.01$ mM) \vs{} added 
salt concentration normalized by the concentration of intrinsic counterions ($2I_s/c_{in}$). Data are for three representative DNA concentrations: $c_{\textrm{DNA}}=0.05$~mg/mL (diamonds, protocol II$1$), $c_{\textrm{DNA}}=0.5$~mg/mL (full triangles, protocol II$1$; open triangles, protocol II$2.1$; open inverse triangles, protocol II$2.2$) and $c_{\textrm{DNA}}=0.8$~mg/mL (circles, protocol II$2.1$). }
\label{fig12}
\end{figure}

Next we address the issue of the respective roles in ionic screening of 
intrinsic DNA counterions and ions from the added salt. First, we examine 
conditions under which the OSF expression for the persistence length $L_p = L_0 
+ a \cdot I_s^{-1}$ is valid. Dielectric data show that the influence of the 
added salt on the persistence length is important as long as the ionic strength 
$I_s$ is sufficiently larger than the concentration of intrinsic counterions. 
Plot in Fig.~\ref{fig12} reveals that this condition reads $2I_s > 0.4 c_{in}$. 
In this limit where the OSF theory applies, the coefficient $a$ should be equal 
0.324 \AA M (assuming MO counterion condensation). However, experimentally, 
different values are found. Measurements of DNA elastic properties as a function 
of ionic strength also yielded the coefficient in the OSF expression, $a = 0.8$ 
\AA M \cite{Baumann97}, which is different from the one expected from the MO 
counterion condensation theory. On the other hand, a magnetic birefringence 
study by Maret and Weill \cite{Maret83} suggested the values of $a$ between 0.25 
and 0.45, therefore indeed close to 0.324. However, the authors fitted their 
data to the $(0.12c_{in} + I_s)^{-1}$ instead to the OSF $I_s^{-1}$ dependence. 
The first term represents the influence of free DNA counterions. It is 
rather obvious that the OSF fit would not yield the MO value of the coefficient 
$a$. Why thus is there a difference between the MO value 
and the measured value of the coefficient in the OSF dependence of the 
persistence length on the ionic strength of the added salt? 

\begin{figure}
\resizebox{0.48\textwidth}{!}{\includegraphics*{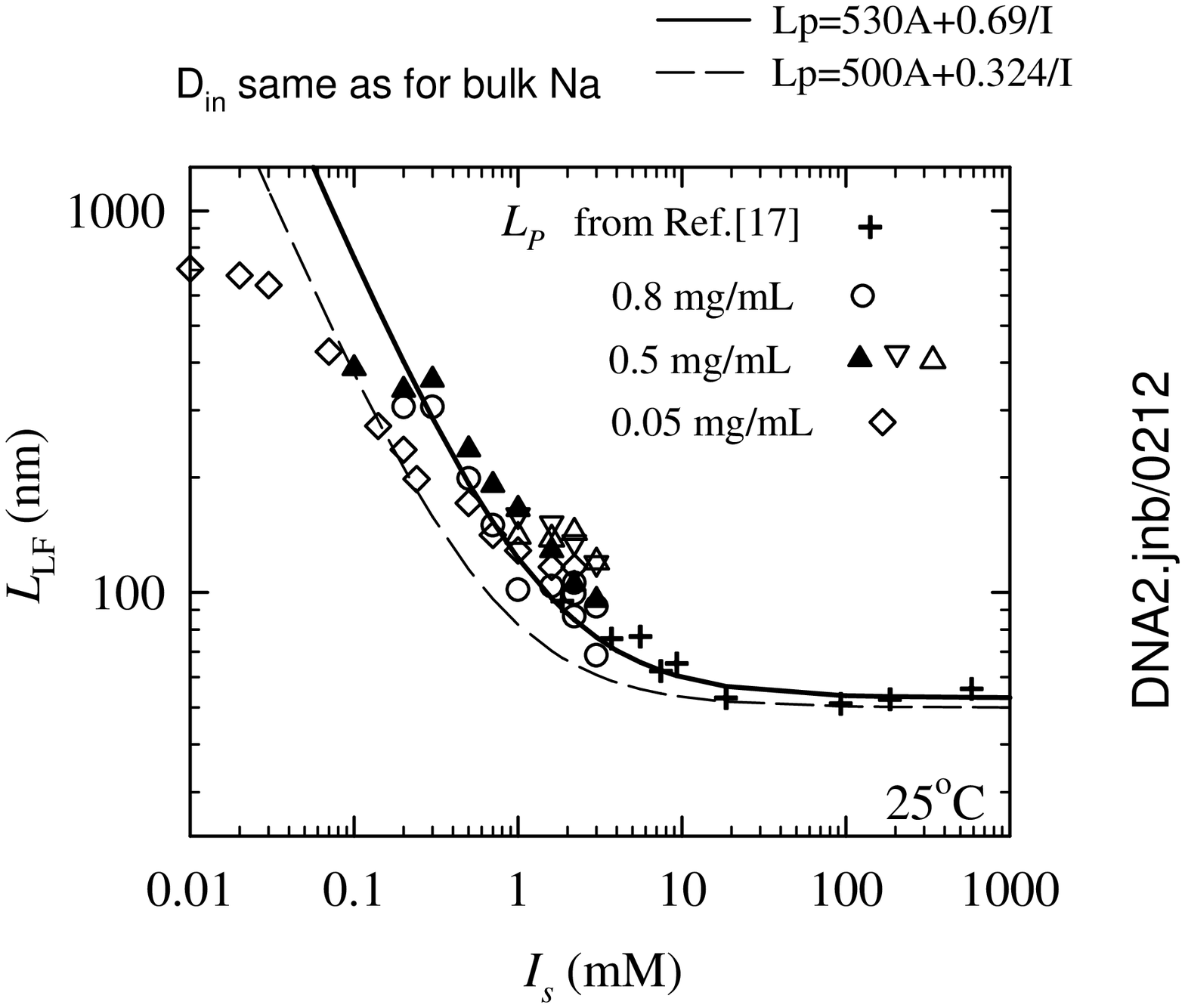}}
%Fig11_0212.JPG
\caption{ Characteristic length of the LF mode, rescaled as  $2.5 \cdot L_{\textrm{LF}} \rightarrow L_{\textrm{LF}}$
($L_{\textrm{LF}}$ data are from Fig.~\ref{fig8}b)) is shown together with data from pulling of DNA (crosses) taken
from Ref.~\cite{Baumann97}. The full line and dashed lines are fits to OSF 
theory with $L_0 = 530$~\AA~and $a = 0.69$ \AA M and with $L_0 = 500$~\AA~and $a = 0.324$ \AA M corresponding to the theoretical OSF prediction with complete MO condensation.} 
\label{fig13}
\end{figure}

In an attempt to reconcile these various values, we have rescaled the 
persistence length from dielectric measurements as $2.5 \cdot L_{\textrm{LF}} 
\rightarrow  L_{\textrm{LF}}$ in order to collapse the behavior from 
dielectric and elastic experiments onto a single curve (Fig.~\ref{fig13}). 
Rescaling is justified since the expression connecting $L_p$ from dielectric properties with the measured mean relaxation time is only valid as a scaling relationship. Numerical factors 
are less straightforward and are essentially unknown. An exact expression 
together with an appropriate numerical coefficient, known as the 
Einstein-Smoluchowski formula, is only known in the dc limit, where $L = 
\sqrt{2D\tau}$.

The fit of this new rescaled version of the $L_{\textrm{LF}}$ to the OSF expression with 
two free parameters now gave $L_0 = 530$ \AA~and $a = 0.69$ \AA M.
The difference between this value of $a$ and the one expected from the MO counterion 
condensation theory signals a different effective linear charge density than the 
one stemming from the MO counterion condensation theory. This difference in the 
measured and theoretically expected value of a coefficient of the OSF fitting 
form for the persistence length would be consistent also with other experiments 
that provide a value for the effective charge density \cite{Podgornik94}.

%%%%%%%%%%%%%%FIGURE%%%%%%%%%%%%%%%%%%%%%%%%
%%%%%%%%%%%%%%FIGURE%%%%%%%%%%%%%%%%%%%%%%%%

%%%%%%%%%%%%%%FIGURE%%%%%%%%%%%%%%%%%%%%%%%%
%%%%%%%%%%%%%%FIGURE%%%%%%%%%%%%%%%%%%%%%%%%

Finally, let us examine the low salt limit ($2I_s < 0.4 c_{in}$) in which we 
expect that the intrinsic counterions become dominant. The characteristic length 
of the LF mode varies with DNA concentration as $L_{\textrm{LF}} \propto 
c_{\textrm{DNA}}^{- 0.29}$, or equivalently as $c_{in}^{-0.29}$. This value of
the exponent would allow us to identify $L_{\textrm{LF}}$ with the
average size of the Gaussian chain composed of correlation blobs scaling as $c_{\textrm{DNA}}^{- 0.25}$, where electrostatic interactions are screened by other chains and counterions, and thus DNA acts as its own salt.

\section{Summary}

All these results make it quite clear that for the dielectric properties studied 
in this contribution the role of free and condensed intrinsic DNA counterions 
can not always be separated. For the HF relaxation we are reasonably 
sure that it is mostly the free counterions that act as the relaxation entities 
in a type of hopping process between the correlated DNA chains in a semidilute 
polyelectrolyte solution. On the contrary, for the LF relaxation process it 
seems to be mostly the condensed counterions that relax along or in close 
proximity to an individual chain in the polyelectrolyte solution, except at 
large enough salt concentrations where at least some of the free counterions seem to 
carry on this role.
 
For the HF relaxation, the experimentally found spatial correlation providing the characteristic  size of the relaxation process is given universally by the polyelectrolyte  semidilute solution correlation length.  The LF relaxation can be characterized by two different correlation  scales. At low salt concentration this correlation length can be identified with the average size of the polyelectrolyte chain, where electrostatic interactions are screened by other chains and counterions, thus where DNA acts as its own salt. At higher salt concentrations, the spatial correlations are provided by the single  chain orientational correlation length, \ie{} the persistence length, that depends on the salt concentration via the OSF mechanism. Both the correlation lengths for the LF relaxation seem to be telling us, that it is the single chain that is responsible for the relaxation process, in which counterions either fluctuate orientationally on the length scale of the persistence length, or they fluctuate  between two ends of the average size of the chain in the solution. 
 
In conclusion, our results demonstrate that there are three fundamental length 
scales that determine the dielectric response of a semidilute DNA solution: the 
average size of the polyelectrolyte chain in the regime in which DNA acts as its own salt, the OSF  salt-dependent persistence length of a single polyelectrolyte chain and the dGPD  semidilute solution correlation length or the mesh size of the polyelectrolyte  solution. While the free DNA counterions can be identified reasonably well as 
the relaxation entities of the HF relaxation mode,  the LF relaxation mode does 
not allow for such a clearcut separation between MO condensed and free 
counterions as relaxation entities. Our data suggest that in fact both of them 
contribute to various extent in different salt and DNA concentration regimes. Finally, our results confirm that although double-stranded DNA at low salt concentrations shows locally exposed hydrophobic cores in a dynamic sense, unzipping of the strands is accomplished only after the denaturation/heating protocol is applied. But even then this unzipping of the two DNA strands is probably at most local and complete separation of the strands at semidilute solutions is never really accomplished. This issue will be addressed in our further studies.

\section*{Acknowledgments}
ICP-AES was performed by M.~Ujevi\'{c} at the Croatian National Institute of Public Health. We thank T.~Ivek, D.~Vurnek and R.~\v{Z}aja for help in the data analysis and electrophoresis measurements. Discussions with D.~Baigl, F.~Livolant and E.~Raspaud are greatly acknowledged. This work was supported by the Croatian Ministry of Science, Education and Sports. Rudi Podgornik would like to acknowledge the financial support of the Agency for 
Research and Development of Slovenia under grant P1-0055(C).

\end{document}